\documentclass[letterpaper]{article}
\usepackage{aaai2026} % preprint/cam-ready, for cam ready, go to the .sty file and undo cas's tracked changes around line 62, also AIES might have a custom sty for us that uses AIES branding/copyright instead of AAAI
\usepackage{times} % DO NOT CHANGE THIS
\usepackage{helvet} % DO NOT CHANGE THIS
\usepackage{courier} % DO NOT CHANGE THIS
\usepackage[hyphens]{url} % DO NOT CHANGE THIS
\usepackage{graphicx} % DO NOT CHANGE THIS
\usepackage{natbib} % DO NOT CHANGE THIS
\usepackage{caption} % DO NOT CHANGE THIS
\usepackage{amsmath}
\usepackage{amsfonts}
\usepackage{booktabs}
\usepackage{array}
\usepackage{xcolor}
\usepackage{float}
\usepackage{colortbl}
\usepackage{subcaption}

\usepackage{newfloat}
\usepackage{listings}
\DeclareCaptionStyle{ruled}{labelfont=normalfont,labelsep=colon,strut=off} % DO NOT CHANGE THIS
\floatstyle{ruled}
\newfloat{listing}{tb}{lst}{}
\floatname{listing}{Listing}
\title{Video Deepfake Abuse: How Company Choices Predictably Shape Misuse Patterns}

\author{
    Max Kamachee\textsuperscript{\rm 1}\equalcontrib,
    Stephen Casper\textsuperscript{\rm 2,\rm 3}\equalcontrib,\\
    Michelle L. Ding\textsuperscript{\rm 4},
    Rui-Jie Yew\textsuperscript{\rm 4},
    Anka Reuel\textsuperscript{\rm 5},
    Stella Biderman\textsuperscript{\rm 6},
    Dylan Hadfield-Menell\textsuperscript{\rm 2}
}
\affiliations{
    \textsuperscript{\rm 1}University of Wisconsin--Madison,
    \textsuperscript{\rm 2}MIT CSAIL,
    \textsuperscript{\rm 3}Harvard Berkman Klein Center,\\
    \textsuperscript{\rm 4}Brown University,
    \textsuperscript{\rm 5}Stanford University,
    \textsuperscript{\rm 6}EleutherAI,\\
    kamachee@wisc.edu, scasper@mit.edu
}

\begin{document}

\maketitle
\begin{abstract}
In 2022, AI image generators crossed a threshold, enabling much more efficient and dynamic production of photorealistic deepfake images than before. This enabled opportunities for creative and positive uses of these models. However, it also enabled unprecedented opportunities for the low-effort creation of AI-generated non-consensual intimate imagery (AIG-NCII), including AI-generated child sexual abuse material (AIG-CSAM). 
Empirically, these harms were principally enabled by a small number of models that were trained on web data with pornographic content, released with open weights, and insufficiently safeguarded. In this paper, we observe ways in which the same patterns are emerging with video generation models in 2025. Specifically, we analyze how a small number of open-weight AI video generation models have become the dominant tools for photorealistic AIG-NCII video generation.
We then analyze the literature on model safeguards and conclude that (1) developers who openly release the weights of capable video generation models without appropriate data curation and/or post-training safeguards foreseeably contribute to mitigatable downstream harm, and (2) model distribution platforms that do not proactively moderate individual misuse or models designed for AIG-NCII foreseeably amplify this harm.
While there are no perfect defenses against AIG-NCII and AIG-CSAM from open-weight AI models, we argue that risk management by model developers and distributors, informed by emerging safeguard techniques, will substantially affect the future ease of creating AIG-NCII and AIG-CSAM with generative AI video tools.
\end{abstract}

% \begin{center}
\noindent \textcolor{red}{Content notice: this paper discusses AI-generated non-consensual intimate imagery and child sexual abuse material.}
% \end{center}

%\tableofcontents

% \newpage  % remove for submission (page limit thing)

\section{Introduction} \label{sec:intro}
In 2022, AI image generation crossed a critical threshold, with models like DALL-E 2 \citep{ramesh2021} and Stable Diffusion 1.x \citep{Rombach2021} making the creation of photorealistic synthetic images much more efficient, dynamic, and accessible than ever before. 
In particular, Stable Diffusion's 1.0's open release in August 2022 made access to these capabilities widely available, allowing large numbers of users to create photorealistic images with minimal technical expertise, very little time, and no specialized data \citep{petsiuk2022human}.
This enabled positive uses of image diffusion systems for recreation, graphics, and art.
However, the accessibility of these new systems enabled unprecedented levels of misuse for AI-generated non-consensual intimate imagery (AIG-NCII), including AI-generated child sexual abuse material (AIG-CSAM).
Since 2022, AIG-NCII and AIG-CSAM images have surged \citep{securityhero2023, iwf2023ai-csam, iwf2024, NCMEC2024, IWF2026}. For example, a report from ActiveFence estimated that the number of ``threads related to the creation of [AIG-]NCII depicting private individuals rose...by 400\%'' between 2022 and 2023 \citep{activefence2023}.
New image diffusion models (principally Stable Diffusion and Flux models, \citealp{schneider2024image, hawkins2025deepfakes}) and the infrastructure built around them (e.g., \citealp{Chai2022}) demonstrated how a small number of models capable of producing realistic NSFW deepfakes can drive misuse patterns (see Section~\ref{sec:history}).

\begin{figure*}[t!]
\centering
\includegraphics[width=1.8\columnwidth]{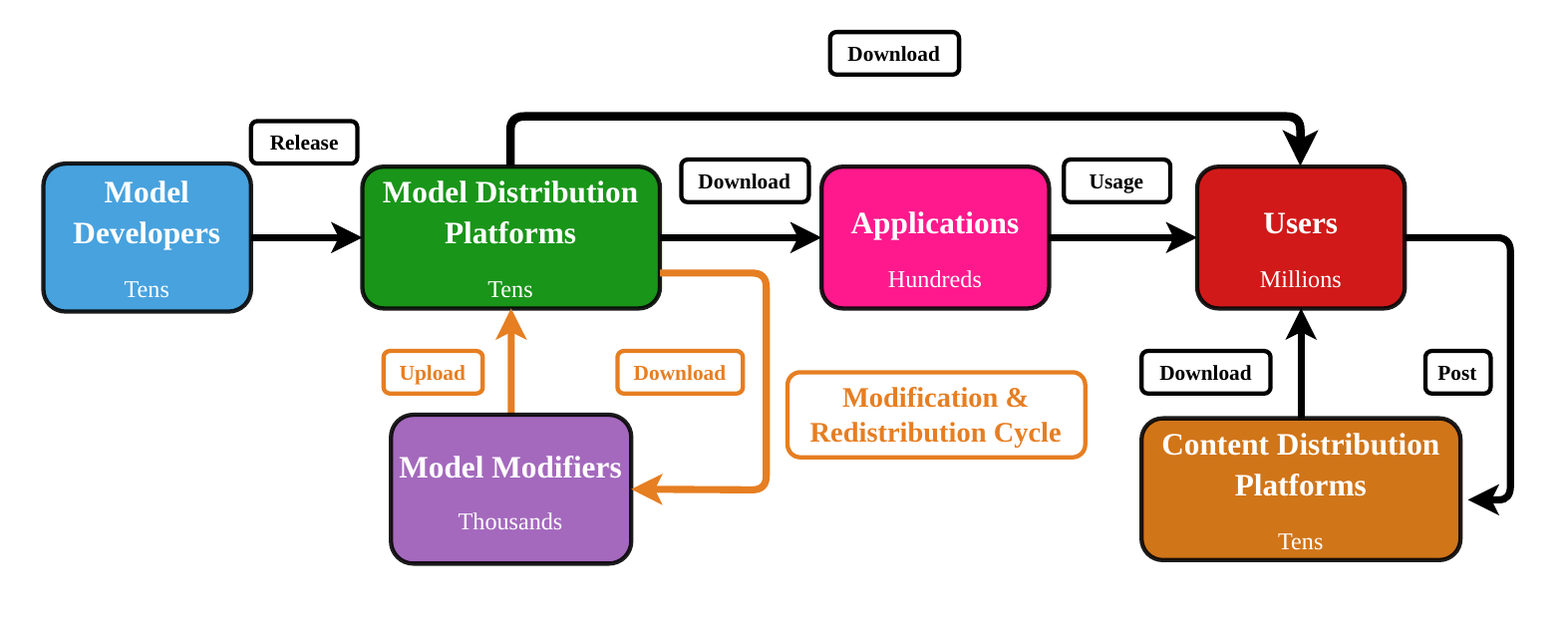}
\caption{\textbf{The supply chain for open-weight AI models capable of creating non-consensual intimate video deepfakes.} Models flow from developers (e.g., Alibaba, Stability AI) through model distribution platforms to modifiers (e.g., Civitai, Hugging Face), who create specialized variants that power user-facing applications (e.g., undressing applications). Individual actors with technical expertise can also directly download models from model distribution platforms and create AIG-NCII locally. The modification and redistribution cycle (highlighted in orange) shows how models with openly available weights can undergo multiple rounds of modification and be re-uploaded to model distribution platforms. Finally, deepfakes can spread via content distribution platforms (e.g., Civitai, X, 4Chan). \textbf{Developers and model distribution platforms serve as critical bottlenecks.} Scale indicators show the rough number of actors at each stage.}
\label{fig:supply_chain}
\end{figure*}

The image generation transformation of 2022 offers precedent for understanding current developments in video generation, where similar capability thresholds are now being crossed. 
Currently, video generation models are undergoing an analogous revolution. Systems such as OpenAI's Sora \citep{openai2024}, Google's Veo \citep{google2025}, and Runway's Gen-4 \citep{runway2025} can be used to produce convincing photorealistic content. 
Meanwhile, as we will discuss in Section~\ref{sec:video}, a small number of open-weight models are emerging as the dominant tools for NSFW video generation, including Wan2.x \citep{wan2025wan}. 
Variants of these models specialized for NSFW content are widely shared across several key online distribution platforms, including Civitai, making these platforms critical gatekeepers for access to these capabilities.

This paper examines how developer and model distribution platform\footnote{This paper uses ``distribution platform'' to refer to model and dataset distribution platforms like Civitai, Hugging Face, and GitHub. These are distinct from further downstream platforms like social media sites or adult sites where AIG-NCII is distributed.} choices can shape patterns of deepfake misuse. By analyzing the history of image generators, AI content shared online, online fora, and academic literature on model safeguards, we present three findings:
\begin{enumerate}
    \item NSFW AI video content online disproportionately stems from a small number of models, including \textit{Wan 2.x}, \textit{Stable Video Diffusion}, \textit{HunyuanVideo}, and \textit{LTX-Video}, whose variants are largely distributed through a small number of online distribution platforms, including \textit{Civitai} (Section~\ref{sec:video}). 
    \item Various technical strategies can impose significant barriers to using these models for AIG-NCII and AIG-CSAM (Section~\ref{sec:mitigations}).
    \item Friction-based mitigations meaningfully reduce harm even when perfect prevention is not possible (Section~\ref{sec:cat}).
\end{enumerate}

\noindent Based on this evidence, we draw two conclusions about how company choices shape downstream misuse patterns:
\begin{enumerate}
    \item AI developers who (a) train photorealistic video generation models without effective filtering of NSFW content, (b) omit sufficient post-training safeguards, and (c) release them with open weights foreseeably contribute to mitigatable downstream harm.
    \item Model distribution platforms that do not proactively monitor models for misuse and take down ones designed for AIG-NCII and AIG-CSAM amplify these harms.
\end{enumerate}

We conclude that the decisions made by AI video generator developers and distributors in the coming months will have the potential to influence misuse patterns into the future, making current attention to risk mitigation important for reducing future harm.

\section{The History of Photorealistic NSFW Image Generation and the 2022 Transformation} \label{sec:history}

\subsection{Methods Prior to 2022}

Prior to the rise of generative AI, convincing synthetic imagery depicting real humans was well-precedented.
Early methods relied on manual image editing tools such as Photoshop, which could enable skilled users to stitch together images of different humans' faces and bodies \citep{eggestein2014fighting}.
In the 2010s, progress in research and development of generative AI added image and video processing models to the toolkit \citep{monaghan2017impact, gieseke2020new, henry2020image}.  
From 2017 onward, there was a notable rise in the availability of both ``AI nudification'' tools (for both images and video) and cultural awareness of them \citep{kobriger2021out, wagner2019word}. 
The gradual rise of tools that could enable AIG-NCII was driven, in part, by the development and release of open software tools from internet communities dedicated to synthetic NSFW content generation \citep{monaghan2017impact, winter2020deepfakes}.

\textbf{Prior to 2022, despite it being possible to generate photorealistic NSFW images and videos of real individuals, tools were limited by skill requirements, a lack of dynamism, inefficiency, and/or inconsistent realism.} 
% Prior to 2022, realistic intimate deepfake images from tools like Photoshop \citep{eggestein2014fighting} and AI systems \citep{gieseke2020new, henry2020image} were precedented.
% However, early techniques were inefficient, lacked widespread accessibility, and/or struggled to consistently and dynamically produce photorealistic deepfakes. 
For example, Photoshop required significant time and experience.
Meanwhile, early generative adversarial networks (GANs) suffered from instability, mode collapse, and limited coherence, producing images that were often easily distinguishable from authentic photographs \citep{saad2024}. 
Prior to 2022, it was particularly challenging for GANs to consistently and dynamically generate realistic images. Models like AttnGAN and DM-GAN performed poorly on standard benchmarks and struggled with complex compositional understanding \citep{xu2017, zhu2019}. The most accessible approaches, such as VQGAN+CLIP combinations, required extensive technical expertise, complex prompt engineering, and produced imagery with characteristic visual artifacts that marked them as clearly synthetic \citep{crowson2022, steinbruck2022}. Even OpenAI's DALL-E 1 diffusion model, while demonstrating impressive conceptual capabilities, operated at low resolution and was not publicly accessible \citep{ramesh2021}.

\subsection{The DALL-E 2 Release}

\textbf{DALL-E 2 ushered in the modern age of dynamic photorealistic image generation.} Released by OpenAI in April 2022, DALL-E 2 marked a leap forward in the ability of AI tools to consistently and dynamically produce approximately photorealistic results at high resolution (1024×1024) from simple text prompts \citep{ramesh2022dalle2}. OpenAI kept the model behind a closed API and reported implementing technical safeguards to prevent misuse, including filtering NSFW training data; internal testing; independent red team evaluations; prompt filtering to block explicit content and celebrity names; and filtering generated images for NSFW content. Furthermore, OpenAI required user registration with email verification and stated intentions; implemented user suspension systems for suspicious activity; and performed continuous backend updates to address emerging issues \citep{dalle2modelcard}. Subsequently, some researchers have found that these safeguards were imperfect, allowing for the production of some NSFW content \citep{yang2024sneakyprompt}. However, to the best of our knowledge, there is no evidence of successful, scalable uses of DALL-E 2 for the creation of individualized AIG-NCII.

\subsection{Stable Diffusion's Release and Downstream Misuse}

\textbf{Stability AI released Stable Diffusion 1.0 without effective mitigations against AIG-NCII misuse.} Stable Diffusion 1.0 was released in August 2022 with open weights. Stability AI released it with a license, a user agreement, beta testing, and an NSFW classifier to block NSFW content. However, these measures proved to offer no substantial barrier to misuse. Usage terms are effectively unenforceable for an open-weight model; the beta testing did not reportedly involve red-teaming for AIG-NCII risks \citep{stable_diffusion_launch_2022}; and the NSFW content classifier could be trivially disabled by users who downloaded the system.\footnote{The Stable Diffusion safety filter was also vulnerable to attacks \citep{Rando2022}.} 
Meanwhile, Stable Diffusion 1.x models were trained on weakly-curated internet data (the LAION-5B dataset), which contained a substantial amount of NSFW content and CSAM, \citep{birhane2311into, thiel2023}.
Meanwhile, Stability AI did not publicly report on any red teaming for NSFW capabilities or AIG-NCII risks. 
As a result of these ineffective safeguards, derivatives of Stable Diffusion 1.x models empirically became effective tools to generate AIG-NCII \citep{schneider2024image, pang2024towards, hawkins2025deepfakes, louk2026deepfakes}.

\textbf{Community-developed tools greatly increased access to AIG-NCII creation in subsequent months.}
Despite the photorealistic image generation power of models like Stable Diffusion 1.x, they were not particularly dynamic or accessible tools by themselves.
For example, initial misuse of them for AIG-NCII was largely focused on generating content of celebrities.
The final key breakthrough toward widespread (mis)use was the introduction of software and applications that greatly increased dynamism and accessibility \citep{wagner2025perpetuating}.  
For example, DreamBooth, released in September 2022, enabled personalization of diffusion models using only three to five reference images \citep{Chai2022}. Using DreamBooth, users could ``teach'' models to associate unique identifiers with specific individuals, then generate novel synthetic content featuring those individuals in any described scenario. 
Meanwhile, other techniques, such as inpainting, outpainting, and stitching,\footnote{Inpainting refers to having a diffusion model construct a customized synthetic version of an image region. Outpainting refers to having a diffusion model synthetically expand an image beyond its original border. Stitching refers to having a diffusion model realistically blend two separate images together (such as inserting one person's head onto another person's body).} allowed users to produce NSFW content of specific subjects from SFW images \citep{shani2021}. 
These innovations meant that a person's social media photos, school yearbook pictures, or family photos were sufficient for creating personalized photorealistic AIG-NCII images, expanding potential targeting from public figures to anyone with minimal digital presence. 
Furthermore, community development quickly produced models further fine-tuned on NSFW data, user-friendly interfaces, pre-configured workflows, and optimized implementations accessible to non-technical users \citep{mink2026unlimited}. 
This decentralized supply chain of these technologies creates a ``malicious technical ecosystem'' \citep{ding2024malicious} that has enabled dozens of ``AI nudifier'' applications, which allow nontechnical users to create AIG-NCII in minutes \citep{gibson2025analyzing, alexios_mantzarlis_ai_2025, williams2025there, ding2026stop}.

\subsection{Misuse Patterns Following Technical Breakthroughs}\label{sec:patterns}

\textbf{A surge in personalized AIG-NCII followed the NSFW image generation breakthroughs of 2022.} 
Within months of Stable Diffusion 1.0's release, dedicated communities formed around generating non-consensual content, sharing techniques for circumventing safety measures, and distributing specialized model configurations optimized for explicit content generation. The vast majority of which depicts women and girls \citep{securityhero2023, laffier2023deepfakes,  my_image_my_choice_deepfake_2024, wagner2025perpetuating}. 
Reports involving generative AI to the National Center for Missing and Exploited Children's CyberTipline surged from 4,700 in 2023 to 67,000 in 2024, a 1,325\% increase \citep{NCMEC2024}.\footnote{These reports include AI-generated CSAM as well as other forms of child sexual exploitation involving generative AI, such as grooming guides and nudify applications.} 
Analysis by Graphika found that 34 synthetic AIG-NCII providers received over 24 million unique visitors to their websites in September 2023, with referral link spam for these services increasing by more than 2,000\% on platforms since the beginning of 2023.
Graphika also found 52 Telegram groups used to access AIG-NCII services containing at least 1 million users as of September 2024 \citep{graphika2023}.
Meanwhile, Activefence reported that ``threads related to the creation of AIG-NCII depicting private individuals increased in the same period by 400\%'' \citep{activefence2023}.
Ultimately, these developments have led to unprecedented patterns of AIG-NCII misuse.
For example, in early 2025, law enforcement identified and made dozens of arrests connected to an international online community with hundreds of members dedicated to AIG-CSAM \citep{europol2025operation}.
% Specialized communities and circumvention tools emerged to exploit new capabilities.
% For example, an analysis by Security Hero studied 15 deepfake creation communities with over 609,464 collective members, finding that one in every three AI generation tools allowed users to create pornographic content \citep{securityhero2023}.

\section{The AI Video Generation Ecosystem in 2025 and 2026} \label{sec:video}

% \subsection{Methodology}

% \begin{quote}
    
% \textit{``Wow this is actually scary…I guess the plus side is that if your nudes ever actually get leaked you can claim ai and then link to this.''}

% - A Reddit user commenting under an NSFW video generated by Wan2.1-VACE-14B, whose creator claimed was ``loosely inspired'' by a famous actress (link omitted due to NSFW content). % https://www.reddit.com/r/unstable_diffusion/comments/1l1lmr1/comment/mvocks4/?context=3
% % Actress is Britt Lower as implied by the OP in comments

% \end{quote}

\begin{quote}
    
\textit{``This is going to really have some wild implications for people who share their photos. Throw a pic up on Instagram, and in ten minutes people can make fairly realistic video clips of you performing sex acts.''}

- A Reddit user commenting under a series of NSFW videos created from an SFW image (link omitted due to NSFW content). % https://www.reddit.com/r/unstable_diffusion/comments/1ofn94l/absolutely_crazy_all_videos_generated_from_the/

\end{quote}

Next, we analyze current patterns in the use of AI video generators for creating NSFW content, observing clear parallels with image models in 2022 and 2023. However, there is limited open research on NSFW video content generation because of ethical barriers, legal barriers, and social taboos. Given the rapid pace of image-generation technology development and the ``underground'' nature of NSFW generation that may not be captured by formal literature, we also analyze online platforms, discussion forums, publicly accessible applications, and user communities.

\subsection{Technical Capabilities and Accessibility}

\textbf{Video generation models in 2025 and 2026 are crossing an analogous capability threshold to what enabled widespread AIG-NCII in 2022.} Current systems, including Google's \textit{Veo}, OpenAI's \textit{Sora}, and Runway's \textit{Gen-3}, can often produce temporally coherent synthetic videos with quality approaching broadcast standards \citep{openai2024, google2025, runway2025}. Motion fidelity and facial consistency have reached levels where synthetic content sometimes becomes difficult to distinguish from authentic footage.\footnote{Subjectively, we attest that a small fraction of the NSFW video deepfakes that we have seen online have appeared to us to be \textit{entirely indistinguishable} from genuine video.}
These capabilities are now being used to create photorealistic AIG-CSAM at scale. 
In January 2026, the Internet Watch Foundation reported a ``26,362\% rise in [discovered] photo-realistic AI videos of child sexual abuse, often including real and recognisable child victims'' between 2024 and 2025 \citep{iwf2026ai}.

\textbf{Model distribution platforms, principally Civitai, facilitate the spread and usage of AI video models for NSFW content.} Model distribution platforms provide infrastructure to help users download models and access detailed workflow documentation \citep{gorwa2024moderating}.
By a large margin, the most prominent distribution platform for distributing models that specialize in NSFW content is Civitai \citep{wagner2025perpetuating, hawkins2025deepfakes, wei2024exploring, civitai2024transparency}, who self reported 446 thousand LoRA adapters trained and 600M images/videos downloaded using their site in 2024.
% Meanwhile, cloud-based services eliminate even the need for local installation by hosting models on remote servers, allowing users to generate content directly through web interfaces without downloading software or managing hardware requirements. 
Community-developed workflows\footnote{\url{https://civitai.com/articles}} further increase the accessibility of NSFW model generation capabilities by providing instructions, workflows, bounties, optimized settings, etc. that enable users to generate custom NSFW videos in minutes \citep{fusionxcivitai2025, ghosh2026marketplace}. 
% These tools provide specialized interfaces for adult content creation, following the same accessibility pattern that enabled widespread AIG-NCII images post-2022. Local deployment options for technically inclined users, cloud-hosted solutions for general consumers, and simplified workflow packages ensure that both computational and technical barriers do little to meaningfully restrict access to AIG-NCII video generation capabilities.

\subsection{Current Patterns and Methods for NSFW Video Generation} \label{sec:patterns}

\textbf{A small number of open-weight models account for the majority of model variants and videos across prominent online NSFW AI video communities} 
To understand current usage patterns, we analyzed model search hits and tags across web communities.
% The full details of our process are outlined in \Cref{app:details}.
For this analysis, we assume that a model's usage for generating NSFW content in general is an informative proxy for its usage in creating AIG-NCII (see Appendix~\ref{app:details} for further discussion). First, we identified 10 popular open-weight video generator model families online.\footnote{We consider a model family to refer to a named and versioned set of models from a single developer, including all releases and derivatives. For example, we considered \texttt{Wan-AI/Wan2.1-T2V-14B}, \texttt{Wan-AI/Wan2.2-TI2V-5B}, and \texttt{Comfy-Org/Wan\_2.2\_ComfyUI\_Repackaged} to all be from the same Wan2.x model family.}
We did this by exhaustively searching for all general-purpose text-to-video, image-to-video, and video-to-video models on Hugging Face from 2024 or later that had more than 10,000 downloads in the past month (as of October 8, 2025).
Next, we identified a set of web platforms that allowed us to search and count AI video models specialized for SFW and NSFW content and AI videos containing SFW and NSFW content. Through web searches and online guides, we explored over 50 websites, pages, applications, Discord servers, and Telegram groups. However, most did not allow us to quantify search hits for models or videos. For example, PixAI lacks search counts, and Tensor.art moderates NSFW content.
Full details of our site selection process are in Appendix~\ref{app:details}.
We ultimately selected six subreddits,
 % -- three with SFW AI video content and three dedicated to NSFW AI video content. 
% r/unstable\_diffusion, r/AIPornhub, and r/sdnsfw
Civitai (the world's most popular community platform for distributing AI models fine-tuned for generating NSFW content \citep{civitai2024transparency}), and a website dedicated specifically to archiving models and content from Civitai to evade moderation. We selected these platforms because they uniquely allowed us to analyze the number of search hits or tags for SFW and NSFW video models or AI videos.
% ncluding and aggregate of subreddits (aggregating r/unstable\_diffusion, r/AIPornhub, and r/sdnsfw), Civitai, and civitaiarchive. 

\begin{figure*}[tp]
    \centering
    \includegraphics[width=\textwidth]{combined_heatmaps.pdf}
    \caption{\textbf{Which open-weight video generation models are the most disproportionately used to create NSFW content online?} We analyze model search hits on subreddits (left), model search hits on Civitai and a Civitai model archive site (middle), and video search hits on Civitai (right). In each analysis, we report the SFW market share, the NSFW market share, and the NSFW/SFW market share ratio. The first two columns of each grid sum to 100\%. Some models, including Wan2.x, stable-video-diffusion, HunyuanVideo, and LTX-Video are disproportionately used to generate NSFW content.}
    \label{fig:all_platforms}
\end{figure*}

To quantify content distribution across platforms, we used a systematic search procedure. 
For subreddits, we searched each model name across three SFW (r/aiArt, r/StableDiffusion, r/aivideo) 
and three NSFW (names omitted)
% (r/unstable\_diffusion, r/AIPornhub, r/sdnsfw) 
subreddits, recording the number of search results returned. For SFW subreddits, we performed these searches using NSFW filters. 
% We aggregated results across the SFW and NSFW set. 
For Civitai and the Civitai model archive website (name omitted), we conducted searches for each model name under two conditions: with the platform's NSFW filter disabled (capturing total content) and enabled (capturing only SFW content), deriving NSFW counts by subtraction. 
For video-specific analysis on Civitai, we used a dual approach. First, we identified NSFW videos by searching with model filters applied and three explicit search terms (``nude,'' ``NSFW,'' ``naked''); second, we captured SFW video content by searching with model filters and an empty search query while enabling the NSFW filter. For each search, we recorded the total number of results reported by the platform's search functionality. From these raw counts, we calculated three metrics for each model-platform combination: SFW market share, NSFW market share, and an NSFW/SFW market share ratio, where values exceeding 1.0 indicate disproportionate usage for NSFW content (Figure~\ref{fig:all_platforms}).

As shown in Figure~\ref{fig:all_platforms}, different models' SFW and NSFW market share differ across platforms. Some models have disproportionately high market share for NSFW content. In particular, this includes Wan2.x, stable-video-diffusion, HunyuanVideo, and LTX-Video. \footnote{For comparison, we also analyzed xAI's \textit{Grok Imagine} model, which is fairly unique among frontier closed-weight AI video generators because its deployers have intentionally enabled its API to produce NSFW video content \citep{maiberg2025grok, rainn2025grok}. We analyzed the prevalence of videos from Grok Imagine on Reddit using the same methodology as with the open models. Despite accounting for 12.7\% of NSFW discussions in our Reddit sample, Grok's NSFW-to-SFW ratio was only 0.58, substantially lower than open-weight models like Wan2.x (1.39) and Stable Video Diffusion (1.04). Note: Our data collection occurred in November 2025. In late December 2025, after our analysis period, Grok's image editing feature was widely misused to create non-consensual intimate imagery, generating up to 7,751 sexualized images per hour at its peak \citep{collier2026grok, lefkowitz2026grok, CCDH2026grok}. This demonstrates that even closed-weight models can face severe misuse when safety controls are insufficient.}

% These findings indicate that \textbf{Wan2.x (Alibaba) is the dominant model for NSFW AI video generation} across the websites we searched. The concentration of NSFW content in a single model family suggests that targeted safety interventions on Wan2.x could have significant impact on reducing misuse.

\section{Mitigating Harms from Open-Weight Models: the Role of Developers and Distributors} \label{sec:mitigations}

\begin{figure*}

\centering
\includegraphics[width=\textwidth]{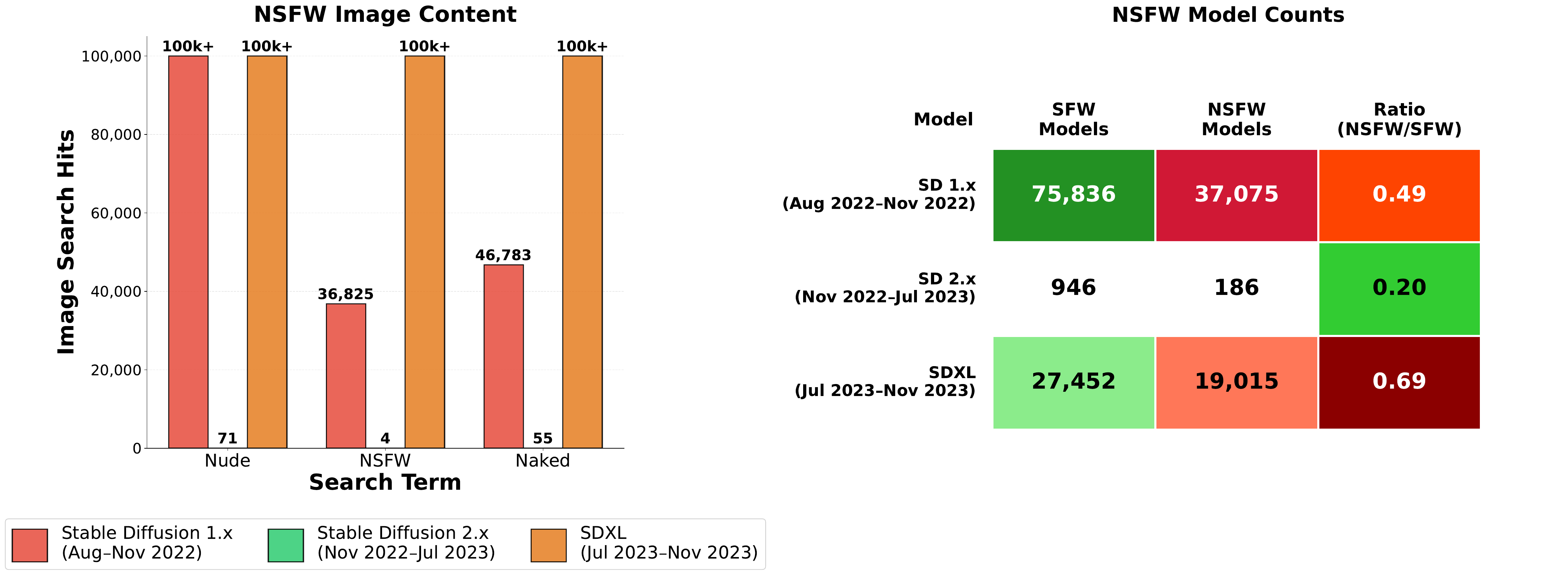}
\caption{\textbf{Stable Diffusion models across three generations demonstrate that filtering NSFW content from training data corresponds to less downstream use for NSFW content.} On Civitai, SD 1.x and SDXL dominate NSFW image searches, with all search terms generating at least tens of thousands of hits. On the other hand, SD 2.x, which were trained on ``an aesthetic subset of the LAION-5B dataset...further filtered to remove adult content using LAION's NSFW filter" \citep{StableDiffV2Release2022}, returns fewer than 100 results per term. Model count analysis shows SD 1.x has 37,075 NSFW-tagged models (ratio 0.49), SDXL has 19,015 NSFW-tagged models (ratio 0.69), and SD 2.x has only 186 NSFW-tagged models (ratio 0.20). SD 2.x's reduction in both NSFW content and models suggests a clear correspondence between training data filtering and a substantial reduction in a model's usage NSFW content.
}
\label{fig:sd_filtering}
\end{figure*}

\subsection{How can Developers Reduce Risks?} \label{sec:developers}

Existing literature on AI safeguards and risk management establishes that several techniques can significantly mitigate harmful uses of open-weight AI models \citep{casper2025open, longpre2024responsible, seger2024open, francois_different_2025, srikumar2024risk, thorn2024safety}. 
Here, in line with \citet{casper2025open}, we focus on risk mitigation techniques for open-weight models that cannot be easily circumvented by users.\footnote{In addition to risk mitigation techniques that we present in Section~\ref{sec:developers}, data provenance methods also serve an important role. While these techniques do not directly mitigate risks, they facilitate a broader understanding of the ecosystem. See \citet{bengio2025international} and \citet{casper2025open}.}
For example, while external safety filters can be valuable to deploy alongside open-weight models (e.g., \citealp{agarwal2025cosmos}), they can be trivially disabled. For instance, Stable Diffusion's safety filter can be bypassed through simple prompt engineering \citep{Rando2022} or disabled entirely by editing the code that runs the model (e.g., setting \texttt{model.safety\_filter = None}).
Crucially, comprehensive AIG-NCII risk mitigation strategies for open models must account for how fine-tuned derivatives of open-weight models, rather than the original models themselves, are empirically a major contributor to misuse \citep{simonovich2024nytheon, casper2025open}.
For example, Civitai hosts thousands of fine-tuned low-rank adapters designed to make video generation models more consistent, dynamic, or specialized for producing NSFW content.

\subsubsection{\textbf{Training Data Curation}}

\textbf{Training data curation is well-precedented, and current evidence suggests that it is a useful intervention for mitigating harmful content generation capabilities.} 
Filtering content from training data related to harmful topics is a well-precedented way to train models with minimal knowledge of such content \citep{nichol2021glide, StableDiffV2Release2022, maini2025safety, o2025deep}.
While the relationship between training data filtering, model capabilities, and resistance to harmful fine-tuning is not yet fully understood in diffusion models, filtering NSFW content from training data is recognized as a key defense against AIG-NCII \citep{NIST_ID012_2024, thorn2024safety}. 
A model trained on NSFW-filtered data can always be taught to generate NSFW content after enough fine-tuning \citep{cretu2025evaluating}, but removing NSFW content from training data substantially increases barriers to misuse by requiring users to fine-tune new NSFW capabilities into a model before it can be misused \citep{o2025deep, nichol2021glide}. 
Dataset curation can also be an important point of harm remediation in machine learning research, as some datasets used in open science contain non-consensually obtained datasets of NSFW content \citep{Cintaqia2025Stop}. 
% find that researchers routinely perpetuate ``image-based sexual abuse'' through the ``nonconsensual creation or distribution of intimate content'' via ``publication, annotation, and open science''.

\textbf{Some approaches to filtering NSFW data can be insufficient to prevent misuse, as suggested by Wan2.x.} While filtering reduces the ease of misuse, it can be insufficient. For example, in the Wan 2.1 technical report, Alibaba reported to ``systematically evaluate and filter inappropriate content based on computed NSFW scores in all training data'' with no additional details provided. However, given the model's documented popularity for NSFW content generation (see Section~\ref{sec:patterns}), this filtering appears not to have been a strong safeguard. Due to a lack of additional public information about how Alibaba filtered NSFW content, it is not possible to critically assess the effectiveness of their specific approach to NSFW data filtering.\footnote{We contacted Alibaba, asking for comment and additional details, but received no reply.} 
However, we can turn to image diffusion models for a notable empirical success story of data filtering.

\textbf{Stable Diffusion models showcase an empirical correspondence between filtering NSFW data and reduced NSFW usage.} 
Between August and October 2022, Stability AI released multiple Stable Diffusion 1.x models \citep{Rombach2021} which were trained relatively indiscriminately on internet data, including NSFW content and CSAM \citep{thiel2023}.
Stable Diffusion 1.x models empirically became the dominant models used to generate AIG-NCII \citep{schneider2024image, pang2024towards, hawkins2025deepfakes, louk2026deepfakes}

Soon after, in November 2022, Stability AI released Stable Diffusion 2.0, reporting that ``Stable Diffusion 2.0 delivers several big improvements and features versus the original V1 release'' \citep{StableDiffV2Release2022} and that
Stable Diffusion 2.0 was trained on ``an aesthetic subset of the LAION-5B dataset...further filtered to remove adult content using LAION's NSFW filter" \citep{StableDiffV2Release2022}. 
Subsequently, the NSFW content generation community has empirically found Stable Diffusion 2.0 to be unhelpful for generating NSFW content \citep{vincent2022stablediffusion}.
For example, we identified various (SFW) Reddit posts reporting ``{We all know by now that SD 2.0 is pretty poor for celebrities, artists, and NSFW content,}''\footnote{\url{https://www.reddit.com/r/StableDiffusion/comments/z4pipp/stable_diffusion_20_is_not_that_bad/}} that Stable Diffusion 2.0 was ``{abandoned because it can not generate any NSFW content,}''\footnote{\url{https://www.reddit.com/r/StableDiffusion/comments/1598xev/comment/jtdu471/?utm_source=share&utm_medium=web3x&utm_name=web3xcss&utm_term=1&utm_content=share_button}} and ``{Well if you’re an overly sensitive prude who has anxiety attacks if you see some [redacted], SD 2.0 is great because it’s so heavily censored.}''\footnote{\url{https://www.reddit.com/r/StableDiffusion/comments/10no2wv/comment/j6a8q5h/?utm_source=share&utm_medium=web3x&utm_name=web3xcss&utm_term=1&utm_content=share_button}}  % redacted text said ``titties'' 
These quotes supplement and help explain what is shown quantitatively in Figure~\ref{fig:sd_filtering}.

Following said backlash, Stability AI appears to have reverted to again training models on NSFW content beginning with SDXL released in July 2023. 
The SDXL technical report did not include details on NSFW content in training data \citep{podell2024sdxl}. 
However, we find in Figure~\ref{fig:sd_filtering} a stark empirical rise in NSFW usage.

Figure~\ref{fig:sd_filtering} quantifies search hits for NSFW content and models on CivitAI with SD 1.x, SD 2.x, and SDXL tags. 
Across three search terms (``nsfw'', ``nude'', and ``naked''), Stable Diffusion 1.x and SDXL models both had dramatically more NSFW content than Stable Diffusion 2.x models: 179,345+ total hits for SD1.x and 300,000+ total hits for SDXL compared to only 128 for SD2.x.\footnote{The ``nsfw'' search for Stable Diffusion 1.x exceeded Civitai's display limit of 100,000 results, so the actual difference is likely substantially larger.} 
Meanwhile, the ratio of NSFW to SFW fine-tuned models is highest for SDXL, followed by SD 1.x, and dramatically lower for SD 2.x, as shown in Figure~\ref{fig:sd_filtering}. 
Taken together, the three generations of models suggest a strong association between NSFW data filtering and reduced NSFW usage, though other factors, such as model architecture changes or community preferences, may also contribute.

\subsubsection{\textbf{Fine-Tuning, Unlearning, and Anti-Tampering}} \label{sec:unlearning}

\textbf{Post-training safeguards can significantly increase barriers to misuse.} 
After a diffusion model is trained, post-training safeguards can offer a second line of defense against misuse of NSFW content generation capabilities.
In particular, developers can utilize ``machine unlearning'' methods to suppress NSFW-generation capabilities. 
In recent years, researchers have introduced and tested a number of approaches to unlearning specific capabilities from diffusion models, including NSFW content \citep{zhang2024defensive, fuchi2024erasing, zhu2024choose, wu2025erasing, zhang2025concept, wu2025unlearning, lu2025concepts}.
To date, the bulk of research on these methods has focused on image diffusion models, but the same algorithms are equally applicable to video diffusion models (e.g., \citealp{liu2024unlearning, xu2025videoeraser, ye2025t2vunlearning}).
While useful, most current unlearning methods are brittle and can be partially circumvented by adversarial prompts or adversarial few-shot fine-tuning \citep{sharma2024unlearning, suriyakumar2024unstable, zhang2024generate, george2025illusion, yu2025forgetme}.
However, there is also emerging research on diffusion model unlearning methods that resist relearning \citep{gao2024meta, li2025towards, abdalla2025gift, li2026towards}.
For example, \citet{abdalla2025gift} introduced an unlearning method that caused diffusion models to resist up to 2,500 steps of fine-tuning on NSFW content while also preserving the model's ability to learn from benign images. 
Overall, research is still ongoing to develop more practical and relearning-resistant unlearning techniques for video diffusion models. 
% Developers must account for, and work to overcome, the limitations of current methods. 
Nonetheless, existing post-training methods for suppressing NSFW generation capabilities offer a barrier to misuse without significantly affecting the SFW capabilities of a diffusion model.

\begin{table*}[h]
\centering
\scriptsize 
\setlength{\tabcolsep}{1mm} 
\begin{tabular}{@{}p{2.7cm}p{2.7cm}p{2.7cm}p{2.7cm}p{2.7cm}p{2.7cm}@{}}
\toprule
& \textbf{NSFW Training Data Curation} & \textbf{Fine-Tuning, Unlearning, \& Anti Tampering} & \textbf{NSFW Capability Evaluations} & \textbf{Staged Deployment} & \textbf{Mention of NCII in Model License} \\ \midrule

\multicolumn{1}{l}{\begin{tabular}{l} \textbf{Wan-AI/Wan2.x} \\ \citet{wan2025wan} \end{tabular}} & \cellcolor[HTML]{edecb9}$<1$ Paragraph 
(One sentence on NSFW data filtration, pg. 6)
& \multicolumn{1}{l}{\cellcolor[HTML]{f0d5d3}No Mention} & \multicolumn{1}{l}{\cellcolor[HTML]{f0d5d3}No Mention} & \multicolumn{1}{l}{\cellcolor[HTML]{f0d5d3}No Mention} & \cellcolor[HTML]{f0d5d3}No Mention \\ \midrule

\multicolumn{1}{l}{\begin{tabular}{l} \textbf{Stable Video Diffusion}\\ \citet{blattmann2023stable} \end{tabular}} & \multicolumn{1}{l}{\cellcolor[HTML]{f0d5d3}No Mention} & \multicolumn{1}{l}{\cellcolor[HTML]{f0d5d3}No Mention} & \multicolumn{1}{l}{\cellcolor[HTML]{f0d5d3}No Mention} & \multicolumn{1}{l}{\cellcolor[HTML]{f0d5d3}No Mention} & \cellcolor[HTML]{cae8ca}$\ge 1$ Paragraph 
(2 sections prohibiting NCII)
\\ \midrule

\multicolumn{1}{l}{\begin{tabular}{l} \textbf{HunyuanVideo} \\ \citet{kong2412hunyuanvideo} \end{tabular}} & \multicolumn{1}{l}{\cellcolor[HTML]{f0d5d3}No Mention} & \multicolumn{1}{l}{\cellcolor[HTML]{f0d5d3}No Mention} & \multicolumn{1}{l}{\cellcolor[HTML]{f0d5d3}No Mention} & \multicolumn{1}{l}{\cellcolor[HTML]{f0d5d3}No Mention} & \cellcolor[HTML]{edecb9}$<1$ Paragraph 
(1 bullet point prohibiting NCII) 
\\ \midrule

\multicolumn{1}{l}{\begin{tabular}{l} \textbf{LTX-Video} \\ \citet{hacohen2024ltx} \end{tabular}} & \multicolumn{1}{l}{\cellcolor[HTML]{f0d5d3}No Mention} & \multicolumn{1}{l}{\cellcolor[HTML]{f0d5d3}No Mention} & \multicolumn{1}{l}{\cellcolor[HTML]{f0d5d3}No Mention} & \multicolumn{1}{l}{\cellcolor[HTML]{f0d5d3}No Mention} & \cellcolor[HTML]{f0d5d3}No Mention \\ \midrule

\multicolumn{1}{l}{\begin{tabular}{l} \textbf{SeedVR2} \\ \citet{wang2025seedvr} \end{tabular}} & \multicolumn{1}{l}{\cellcolor[HTML]{f0d5d3}No Mention} & \multicolumn{1}{l}{\cellcolor[HTML]{f0d5d3}No Mention} & \multicolumn{1}{l}{\cellcolor[HTML]{f0d5d3}No Mention} & \multicolumn{1}{l}{\cellcolor[HTML]{f0d5d3}No Mention} & \cellcolor[HTML]{f0d5d3}No Mention \\ \midrule

\multicolumn{1}{l}{\begin{tabular}{l} \textbf{CogVideoX} \\ \citet{yang2024cogvideox} \end{tabular}} & \cellcolor[HTML]{edecb9}$<1$ Paragraph 
(Said in email correspondence that an unspecified procedure for NSFW filtration was used.)
& \multicolumn{1}{l}{\cellcolor[HTML]{f0d5d3}No Mention} & \cellcolor[HTML]{edecb9}$<1$ Paragraph (Said in email correspondence that an unspecified procedure for NSFW capability evals was used.)
& \multicolumn{1}{l}{\cellcolor[HTML]{f0d5d3}No Mention} & \cellcolor[HTML]{f0d5d3}No Mention \\ \midrule

\multicolumn{1}{l}{\begin{tabular}{l} \textbf{AnimateDiff-Lightning} \\ \citet{lin2024animatediff} \end{tabular}} & \multicolumn{1}{l}{\cellcolor[HTML]{f0d5d3}No Mention} & \multicolumn{1}{l}{\cellcolor[HTML]{f0d5d3}No Mention} & \multicolumn{1}{l}{\cellcolor[HTML]{f0d5d3}No Mention} & \multicolumn{1}{l}{\cellcolor[HTML]{f0d5d3}No Mention} & \cellcolor[HTML]{edecb9}$<1$ Paragraph (2 bullet points prohibiting NCII) \\ \midrule

\multicolumn{1}{l}{\begin{tabular}{l} \textbf{Stable Virtual Camera} \\ \citet{zhou2025stable} \end{tabular}} & \multicolumn{1}{l}{\cellcolor[HTML]{f0d5d3}No Mention} & \multicolumn{1}{l}{\cellcolor[HTML]{f0d5d3}No Mention} & \multicolumn{1}{l}{\cellcolor[HTML]{f0d5d3}No Mention} & \multicolumn{1}{l}{\cellcolor[HTML]{f0d5d3}No Mention} & \cellcolor[HTML]{cae8ca}$\ge 1$ Paragraph (2 sections prohibiting NCII) \\ \midrule

\multicolumn{1}{l}{\begin{tabular}{l} \textbf{Cosmos} \\ \citet{agarwal2025cosmos} \end{tabular}} & \multicolumn{1}{l}{\cellcolor[HTML]{f0d5d3}No Mention} & \multicolumn{1}{l}{\cellcolor[HTML]{f0d5d3}No Mention} & \cellcolor[HTML]{cae8ca}$\ge 1$ Paragraph (Red-teaming for sexual content and CSAM, pg. 55) & \multicolumn{1}{l}{\cellcolor[HTML]{f0d5d3}No Mention} & \cellcolor[HTML]{edecb9}$<1$ Paragraph (1 sentence prohibiting NCII) \\ \midrule

\multicolumn{1}{l}{\begin{tabular}{l} \textbf{Mochi 1} \\ \citet{genmo2025mochi1} \end{tabular}} & \multicolumn{1}{l}{\cellcolor[HTML]{f0d5d3}No Mention} & \multicolumn{1}{l}{\cellcolor[HTML]{f0d5d3}No Mention} & \multicolumn{1}{l}{\cellcolor[HTML]{f0d5d3}No Mention} & \multicolumn{1}{l}{\cellcolor[HTML]{f0d5d3}No Mention} & \cellcolor[HTML]{f0d5d3}No Mention \\ \bottomrule
\end{tabular}%
% }
\caption{\textbf{Popular open-weight AI video generator developers rarely report on AIG-NCII risks and mitigations in model technical reports.} When risks or mitigations related to AIG-NCII or AIG-CSAM are discussed, it is often with very little detail.}
\label{tab:reporting}
\end{table*}

\subsubsection{\textbf{Evaluations}}

\textbf{Evaluating open-weight models for misuse potential, particularly under fine-tuning threats, is needed to meaningfully assess real-world misuse risks.}
Evaluations (both internal and external) of frontier AI systems' capabilities are increasingly recognized as a key pillar of AI risk management frameworks.
% For example, the EU AI Act codes of practice require that frontier AI system developers conduct evaluations of frontier models as needed to assess systemic risks using appropriate access and elicitation techniques \citep{EU_GPAI_Code_2025}.
Safety evaluations of frontier systems are most commonly conducted using input-output access in which evaluators give models queries and analyze their responses. 
For closed weight models, input-output evaluations emulate the real-world misuse threats that models will be exposed to. 
However, for open-weight models, input-output evaluations do not fully account for the risks posed by simple downstream modifications \citep{casper2024black, che2025model, casper2025open}. 
It is intuitive that evaluating the safety of open-weight models requires adversarial fine-tuning, and some developers (e.g., \citealp{wallace2025estimating}) do so. However, there is limited precedent for open-weight models to be evaluated under harmful fine-tuning attacks.
For example, while some of the developers we analyzed in Section~\ref{sec:video} report on evaluations for NSFW capabilities (see Table~\ref{tab:reporting}, zero mention evaluations under fine-tuning.

\subsubsection{\textbf{Staged Deployment}}

\textbf{Staged release strategies enable monitoring and patches prior to an open-weight release.} 
Model deployment is not a binary between fully closed and fully open.
Different deployment strategies offer different trade-offs between openness and security.
When a company has the resources to implement them, these options allow model developers to gradually increase access to a model before choosing to fully release it with open weights \citep{solaiman2023gradient}.
For example, beta testing \citep{ujkani2025} and fine-tuning APIs \citep{wu2024finetunebench} are common approaches by which developers can study and patch risks before a full open-weight release.
These strategies allow developers to monitor how real users interact with their systems, identify emerging misuse patterns, and refine safety measures before a full open release \citep{casper2025open}.
For example, observations following a staged deployment could help the developer identify a need for additional mitigations, such as unlearning (Section~\ref{sec:unlearning}).

\begin{figure*}[tp]
    \centering
    \includegraphics[width=0.9\textwidth]{model_search_results.pdf}
    \caption{\textbf{Model search results across major distribution platforms for ``dog whistle'' terms often associated with AIG-NCII (May 2026).} One of these terms appearing in a model's name is not a clear sign that it specializes in photorealistic AIG-NCII/AIG-CSAM. However, this analysis suggests that platforms may host a substantial number of models designed for and capable of AIG-NCII/AIG-CSAM generation, particularly Civitai.}
    \label{fig:model_search}
\end{figure*}

\subsection{Developers Rarely Report on Mitigations} \label{sec:reporting}

\textbf{Most open-weight model developers do not report on efforts to safeguard their models against AIG-NCII misuse.}
Above in Section~\ref{sec:patterns}, we observe that some models, including Wan2.1, stable-video-diffusion, HunyuanVideo, and LTX-Video are disproportionately used for NSFW content. 
To better understand why this might be the case, we analyze technical reports from developers for mentions of the mitigation strategies discussed above in Section~\ref{sec:developers}.\footnote{Our analysis was limited to reports available online in English. However, some of the developers in Table~\ref{tab:reporting} are based in countries where English is not the official language. Further details could be available in non-English languages.}
We summarize results for the same ten models as in Section~\ref{sec:patterns} in Table~\ref{tab:reporting}.
Developers consistently report very little about AIG-NCII risks or risk mitigations.
And even when information is presented, it is often with very little detail.

\textbf{The lack of reporting from developers makes it challenging to empirically study practical AIG-NCII risk mitigations for open-weight video models.}
While prior research discussed above in Section~\ref{sec:developers} has established the effectiveness of some safeguards for open-weight models, the lack of reporting from developers on whether/how these techniques were used makes it difficult to assess their usefulness and practice.
We reached out to all 10 developers for comment and additional information, receiving only one brief reply from the developers of CogVideoX, saying that an unspecified quantity of NSFW data filtering and NSFW capability evaluations.
Overall, based on usage patterns observed in Section~\ref{sec:video} and Figure~\ref{fig:all_platforms}, there is evidence that some models were ineffectively safeguarded against NSFW capabilities,
particularly Wan2.1, stable-video-diffusion, HunyuanVideo, and LTX-Video (see Figure~\ref{fig:all_platforms}).
This, combined with sparse reporting, may be indicative of an industry pattern of omitting established mitigations for misuses of NSFW AI-generated video content. 
However, due to insufficient information, it is not possible to understand current industry practices with precision.

\subsection{How Can Model Distribution Platforms Reduce Risks?} \label{sec:distributors}

\textbf{Model distribution platforms are key to the accessibility of harmful model capabilities.}
Thus far, we have primarily focused on video model developers because development is a key point of influence and leverage in the AI ecosystem \citep{williams2025regulating, arcila2025ai}.
% However, other infrastructure for NCII-capable models and potential NCII content are critical to the supply chain for AI-generated NCII.
However, while developers are responsible for creating models, distribution platforms are principally responsible for their accessibility \citep{gorwa2024moderating}.

\textbf{Distribution platforms can amplify misuse by hosting and distributing models designed for AIG-NCII and falling short on content moderation.} 
% Model distribution platforms are a critical node in the supply chain for AIG-NCII-capable models.
Many early image-generation tools for generating AIG-NCII were hosted and distributed on GitHub, one of the most popular developer platforms in the world. In 2023, the National Center on Sexual Exploitation reported that GitHub had been hosting open-source repositories that enable AIG-NCII generation. This included DeepFaceLab, which ``contains direct links to the most prolific sexual deepfake abuse website in the United States: Mr.DeepFakes,'' as well as repositories for DeepNude and Unstable Diffusion \cite{patrick_trueman_esq_github_2023}.

In the past several years, generative AI model distribution has increasingly shifted to Civitai and Hugging Face \citep{wei2024exploring, hashim2025, wagner2025perpetuating}. In May 2025, following new regulations and pressure from payment processors, Civitai banned models designed to recreate real individuals' likenesses, which, according to 404 Media, dealt a ``major blow to the nonconsensual AI porn ecosystem'' \cite{civitai404media, civitai404media1}. 
However, the deepfake community subsequently organized a coordinated effort to archive over 5,000 of these models and re-upload them to Hugging Face \citep{maiberg2025}. 
Recently, Hugging Face has been found to host models specifically designed for producing AIG-NCII of celebrities, including individuals who were minors when source material was created \citep{hashim2025, maiberg2025}. 
These models violated Hugging Face's stated content policy prohibiting ``sexual content created without explicit consent'' and ``underage nudity or any sexual content involving minors'' \citep{hfcontentpolicy2025}. Nonetheless, \citet{maiberg2025} found that Hugging Face was slow to take these models down, allowing some to remain accessible for nearly five months despite multiple requests. Even after removal events, some violating models remained accessible \citep{hashim2025, ghosh2026marketplace}. 

We searched major model distribution platforms for ``dog whistle'' terms commonly associated with AIG-NCII and AIG-CSAM production (`realistic,' `teen,' `young,' `girl,' `petite', `nudif', `undress'), finding over 29,000 models across Civitai and Hugging Face (Figure~\ref{fig:model_search}).\footnote{For Civitai, we filtered for NSFW generators; for Hugging Face, we filtered for image and video generators. Tensor.art and PixAI were excluded due to active moderation and lack of search counts, respectively.} While individual terms don't confirm specialization in AIG-NCII/AIG-CSAM, this suggests platforms host substantial numbers of such models, particularly Civitai.

\section{Discussion}\label{sec:discussion}

\subsection{Are AIG-NCII Harms from Open Video Generators Unmitigatable?} \label{sec:cat}

In this paper, we have argued that risk management practices related to data curation, safeguards, evaluations, release, and platform moderation (Section~\ref{sec:mitigations}) will have a significant effect on the production of AIG-NCII. 
But are mitigations effective if they are not used universally?
A possible argument against prioritizing risk mitigation is that the value of one company's efforts can be undercut by another company that does not prioritize risk mitigation.
We term this the ``cat-out-of-the-bag'' argument.
With models such as Wan2.x and Stable Video Diffusion already being released and distributed, is the cat out of the bag? 
Does their existence (or the future existence of other models like them) nullify the value of future models and distribution platforms being safeguarded to mitigate AIG-NCII? 
We argue it does not. 

\textbf{The ``cat-out-of-the-bag" argument overlooks harm reduction principles and incorrectly characterizes technology adoption dynamics.} The `cat-out-of-the-bag' reasoning is not in line with established approaches to risk management and misidentifies how technology proliferates. 
Mitigations matter, especially in a domain where a substantial fraction of perpetrators are teenagers.
For example, digital piracy provides one clear parallel: despite the theoretical availability of pirated content online, coordinated enforcement efforts by search engines, authorities, and companies to reduce the accessibility of pirated content online have had a measurable impact. 
For example, \citet{smith2023piracy} concluded that removing links to pirated content from search results increases legal content sales by 11-14\%, and \citep{danaher2017} concluded that search engines that demote piracy sites in results shift user behavior toward legal consumption. 
Even though determined users can still find pirated content with sufficient effort, increasing barriers to accessing it meaningfully reduces piracy by increasing search costs and directing casual users toward legitimate alternatives.
Reducing the accessibility of harmful generative AI capabilities may be particularly helpful for reducing AIG-NCII given that many perpetrators are low-skill minors \citep{alexander2025deepfake}.

\textbf{Risk mitigation operates through aggregate effects, creating systemic benefits even when circumvention remains possible.} Safety measures implemented by major platforms and developers create systemic benefits, including reducing total explicit content volume by avoiding accidental harms, preventing unsophisticated misuse, deterring costly circumventions of safeguards, preventing normalization by keeping harmful applications marginalized rather than mainstream, and enabling legal frameworks that can target deliberate safety circumvention \citep{ec2019}.

\textbf{Technology adoption follows network effects that amplify early choices made by major actors.} The first widely available implementations establish user expectations, development priorities, and community norms that influence subsequent development. 
For example, the LAION dataset was introduced in March 2022 \citep{laion2022laion5b}, was found to contain CSAM in December 2023 \citep{thiel2023}, and was modified to mitigate CSAM content in August 2024 \citep{laion2024relaion5b}.
Nonetheless, prior to August 2024, the original LAION dataset was widely used to train generative systems, including Stable Diffusion models \citep{Salvaggio2024}, that are still commonly used today.
When major developers and platforms implement robust safety measures, they also create incentives for compatible approaches across the ecosystem. Conversely, early releases without safety measures normalize applications and create communities focused on circumvention rather than beneficial use.
As such, the ``cat-out-of-the-bag'' argument offers a pretext to race to the bottom.

\subsection{Policy Influences Company Incentives}

This paper has focused on how the choices made by developers and model distribution platforms will affect downstream misuse patterns for AI video generators. 
However, it is key, especially for policy audiences, to recognize that these choices are not made in a vacuum, but amidst external commercial, regulatory, and liability incentives.  
Diffusion model generators and distribution platforms have commercial incentives to deploy and share models as quickly as possible, but regulatory requirements, liability, and transparency can have a significant counteractive effect to incentivize risk mitigation \citep{bommasani_considerations_2024, smith2024liability, ramakrishnan2024tort, williams2025regulating, arcila2025ai}. 
When considering policies related to managing AI risks-- especially from open-weight models, policymakers should recognize that some types of risk management techniques, such as staged deployment, may be more onerous for small developers and academics than for larger companies.

\textbf{Some laws on AIG-NCII have been written to regulate content and models.} Some regulatory approaches have aimed to influence the AIG-NCII ecosystem by targeting downstream non-consensually distributed AIG-NCII, such as the federal TAKE IT DOWN Act in the USA \citep{sen_cruz_ted_r-tx_take_2023}.
Other approaches can also focus on models themselves. 
For example, Arkansas Act 827, enacted April 2025, states that ``the Attorney General may institute a civil action on behalf of the state against a provider or developer of image generation technology that was used to create deepfake visual material'' if ``the provider or developer of the image generation technology did not have reasonable safeguards in place to protect against the generation of deepfake visual material'' \cite{state_of_arkansas_act_2025}. 

\textbf{Transparency requirements invite external scrutiny.} 
Another major factor in the incentives of developers and model distributors is transparency \cite{widder2022limits}. 
Detailed information about risk management measures gives external stakeholders the ability to study their mitigations.
A low-transparency state of affairs, as illustrated in Table~\ref{tab:reporting}, can be expected by default because selective transparency helps companies avoid unwanted forms of scrutiny \citep{ananny2018seeing}.
For example, the \textit{lack of} information about AIG-NCII risk mitigations from developers (see Table~\ref{tab:reporting}) impedes work like ours from directly identifying specific links between company choices and downstream harms.
Absent external incentives, a low transparency regime will tend to self-reinforce by inviting increased levels of scrutiny and criticism of developers who voluntarily choose to be more transparent.

\subsection{Weighing Competing Considerations}

\textbf{Navigating the benefits and risks of open-weight models:} 
As we have discussed, models with openly available parameters are a key factor and key driver behind the creation of AIG-NCII and AIG-CSAM \citep{iwf2023ai-csam, widder2022limits}. 
However, this does not imply open-weight models are net harmful. 
They are also key for positive uses, including the de-concentration of power and the facilitation of open research on safety \citep{kapoor_societal_2024, bommasani_considerations_2024, francois_different_2025}. 
More pragmatically, powerful open-weight models are increasingly influential \citep{bhandari2025forecasting} and likely inevitable.
In recognition of this, we have emphasized a harm mitigation approach for safeguarding open-weight models in Section~\ref{sec:mitigations}, arguing that substantial reduction of AIG-NCII harms can be achieved without prohibiting powerful open-weight models.

\textbf{Balancing the benefits and risks of NSFW AI Content:} 
AIG-NCII is a form of image-based sexual abuse that results in psychological, financial, and reputational harm, disproportionately borne by women and girls 
% as well as a gendered ``chilling effect''
\cite{american_sunlight_project_deepfake_2024, henry_ajder_state_2019, flynn_deepfakes_2022}.
Prior works have argued that the current NSFW generative AI ecosystem is characterized by non-consensual practices \cite{gibson2025analyzing, ding2024malicious}.
However, AI models that produce NSFW content 
% of consenting adults 
could also have applications 
% that promote digital intimacy and sex positivity 
based on consensual and ethical use
\citep{geeng2025co}. Work to build infrastructure for more positive uses of NSFW AI content between adults would require consent-focused data collection, conscientious deployment, and concerted effort to reduce AIG-NCII and AIG-CSAM risks \citep{geeng2025co, Cintaqia2025Stop}.

\textbf{Considerations related to free expression:} A technical reality of managing AI deepfake risks is that the same underlying model capabilities needed to produce consensual content are either identical or very similar to the model capabilities needed to produce  AIG-NCII and AIG-CSAM \citep{cretu2025evaluating}. This poses tradeoffs between efforts to reduce harmful misuse and preserve access for consensual uses. In some jurisdictions, consensual nudification is likely to be recognized as a protected form of speech (e.g., \citealp{dugan2003regulating}). 

\subsection{Limitations}

\textbf{Our analysis focused on English-language platforms, which may not represent global AIG-NCII production patterns.} Deepfake AI pornography is differently prevalent and differently viewed across cultures \citep{umbach2024non}. For example, in one study, \citet{securityhero2023} found a striking prevalence of deepfakes targeting South Korean singers and actresses (53\% of targets identified). Meanwhile, cultural values shape how societies view and regulate this technology \citep{krkic2025cultural}. Our framing and findings may not easily generalize to other linguistic or cultural contexts.

\textbf{Studying the NSFW content generation ecosystem remains challenging, especially in a low-transparency regime.} 
Our findings are necessarily based on observable reporting, usage patterns, and analysis of discussions in online forums. For example, with little reporting from developers about how they mitigate risks, we have not been able to directly analyze the relationship between mitigations and downstream harms. 
Future technical work should continue to study the NSFW AI image and video generation ecosystem toward a better understanding of the relationship between upstream decisions and downstream harms.
However, absent public pressure or regulatory incentives for companies to meaningfully report about risk assessments and risk mitigation, the empirical study of AIG-NCII and AIG-CSAM harms will continue to be challenging.

% \section*{Acknowledgments}

% We are grateful to Markus Anderljung for discussion and feedback.

% \bibliographystyle{apalike}
\bibliography{references}

@inproceedings{Chai2022,
  author       = {Chai, M. and et al.},
  title        = {DreamBooth: Fine-Tuning Text-to-Image Diffusion Models for Subject-Driven Generation},
  booktitle    = {Advances in Neural Information Processing Systems},
  year         = {2022},
  note         = {arXiv:2208.12242},
}

@inproceedings{Cintaqia2025Stop,
  title={Stop the Nonconsensual Use of Nude Images in Research},
  author={Cintaqia, Princessa and Arya, Arshia and Redmiles, Elissa and Kumar, Deepak and McDonald, Allison and Qin, Lucy},
  booktitle={Advances in Neural Information Processing Systems (NeurIPS) 2025},
  year={2025},
  note={Poster 121948},
  url={https://neurips.cc/virtual/2025/poster/121948},
}

@article{american_sunlight_project_deepfake_2024,
	title = {Deepfake {Pornography} {Targeting} {Members} of {Congress}},
	url = {https://www.americansunlight.org/updates/deepfake-pornography-targeting-members-of-congress},
	urldate = {2025-02-27},
	author = {{American Sunlight Project}},
	month = dec,
	year = {2024},
}

@article{bengio2025international,
  title={International ai safety report},
  author={Bengio, Yoshua and Mindermann, S{\"o}ren and Privitera, Daniel and Besiroglu, Tamay and Bommasani, Rishi and Casper, Stephen and Choi, Yejin and Fox, Philip and Garfinkel, Ben and Goldfarb, Danielle and others},
  journal={arXiv preprint arXiv:2501.17805},
  year={2025}
}

@techreport{henry_ajder_state_2019,
	title = {The {State} of {Deepfakes}: {Landscape}, {Threats}, and {Impact}},
	url = {https://regmedia.co.uk/2019/10/08/deepfake_report.pdf},
	institution = {Deeptrace},
	author = {{Henry Ajder} and {Giorgio Patrini} and {Francesco Cavalli} and {Laurence Cullen}},
	month = sep,
	year = {2019},
}

@misc{state_of_arkansas_act_2025,
	title = {Act 827},
	url = {https://arkleg.state.ar.us/Bills/Detail?id=HB1529&ddBienniumSession=2025%2F2025R&Search=},
	author = {{State of Arkansas}},
	month = apr,
	year = {2025},
}

@article{civitai404media,
    author = {Emanuel Maiberg},
    title = {Civitai Ban of Real People Content Deals Major Blow to the Nonconsensual AI Porn Ecosystem},
    journal = {404 Media},
    year = 2025,
    howpublished = {Retrieved from \url{https://www.404media.co/civitai-ban-of-real-people-content-deals-major-blow-to-the-nonconsensual-ai-porn-ecosystem/}}
}

@article{civitai404media1,
    author = {Emanuel Maiberg},
    title = {Civitai, Site Used to Generate AI Porn, Cut Off by Credit Card Processor},
    journal = {404 Media},
    year = 2025,
    howpublished = {Retrieved from \url{https://www.404media.co/civitai-site-used-to-generate-ai-porn-cut-off-by-credit-card-processor/}}
}

@article{saad2024,
    author = {Saad, M and et. al.},
    title = {A survey on training challenges in generative adversarial networks for biomedical image analysis},
    journal = {Artificial Intelligence Review},
    year = {2024},
    howpublished = {Retrieved from \url{https://doi.org/10.1007/s10462-023-10624-y}}
}

@misc{EC2019,
  author       = {{European Commission}},
  title        = {Ethics Guidelines for Trustworthy AI},
  year         = {2019},
  howpublished = {Retrieved from \url{https://digital-strategy.ec.europa.eu/en/library/ethics-guidelines-trustworthy-ai}},
}

@misc{StableDiffV2Release2022,
    author = {StabilityAI},
    title = {Stable Diffusion 2.0 Release},
    year = {2022},
    howpublished = {Retreived from \url{https://stability.ai/news/stable-diffusion-v2-release}}
}

@misc{Google2025,
  author       = {Google},
  title        = {Veo 3 Model Card},
  year         = {2025},
}

@article{ding2024malicious,
    author = {Ding, Michelle and Suresh, Harini},
    title = {The Malicious Technical Ecosystem: Exposing Limitations in Technical Governance of AI-Generated Non-Consensual Intimate Images of Adults},
    journal = {Arxiv},
    year = {2025},
    howpublished = {Retrieved from \url{https://arxiv.org/pdf/2504.17663}}
}

@techreport{NCMEC2024,
  author       = {{National Center for Missing \& Exploited Children}},
  title        = {CyberTipline Report},
  year         = {2024},
  institution = {NCMEC},
  howpublished = {Retrieved from \url{https://ncmec.org/gethelpnow/cybertipline/cybertiplinedata}},
}

@article{ananny2018seeing,
  title={Seeing without knowing: Limitations of the transparency ideal and its application to algorithmic accountability},
  author={Ananny, Mike and Crawford, Kate},
  journal={new media \& society},
  volume={20},
  number={3},
  pages={973--989},
  year={2018},
  publisher={SAGE Publications Sage UK: London, England}
}

@inproceedings{xu2025videoeraser,
  title={VideoEraser: Concept Erasure in Text-to-Video Diffusion Models},
  author={Xu, Naen and Zhang, Jinghuai and Li, Changjiang and Chen, Zhi and Zhou, Chunyi and Li, Qingming and Du, Tianyu and Ji, Shouling},
  booktitle={Proceedings of the 2025 Conference on Empirical Methods in Natural Language Processing},
  pages={5965--5994},
  year={2025}
}

@techreport{ramakrishnan2024tort,
  author = {Ramakrishnan, Ketan and Smith, Gregory and Downey, Conor},
  title = {U.S. Tort Liability for Large-Scale Artificial Intelligence Damages: A Primer for Developers and Policymakers},
  institution = {RAND Corporation},
  year = {2024},
  number = {RR-A3084-1},
  doi = {10.7249/RRA3084-1},
  url = {https://www.rand.org/pubs/research_reports/RRA3084-1.html}
}

@article{ye2025t2vunlearning,
  title={T2VUnlearning: A Concept Erasing Method for Text-to-Video Diffusion Models},
  author={Ye, Xiaoyu and Cheng, Songjie and Wang, Yongtao and Xiong, Yajiao and Li, Yishen},
  journal={arXiv preprint arXiv:2505.17550},
  year={2025}
}

@techreport{smith2024liability,
  title = {Liability for Harms from {AI} Systems: The Application of {U.S.} Tort Law and Liability to Harms from Artificial Intelligence Systems},
  author = {Smith, Gregory and Stanley, Karlyn D. and Marcinek, Krystyna and Cormarie, Paul and Gunashekar, Salil},
  institution = {RAND Corporation},
  type = {Research Report},
  number = {RR-A3243-4},
  year = {2024},
  address = {Santa Monica, CA},
  url = {https://www.rand.org/pubs/research_reports/RRA3243-4.html},
  note = {Accessed November 14, 2025}
}

@article{alexander2025deepfake,
  title={Deepfake Cyberbullying: The Psychological Toll on Students and Institutional Challenges of AI-Driven Harassment},
  author={Alexander, Sergio},
  journal={The Clearing House: A Journal of Educational Strategies, Issues and Ideas},
  volume={98},
  number={2},
  pages={36--50},
  year={2025},
  publisher={Taylor \& Francis}
}

@inproceedings{widder2022limits,
  title={Limits and possibilities for “ethical ai” in open source: A study of deepfakes},
  author={Widder, David Gray and Nafus, Dawn and Dabbish, Laura and Herbsleb, James},
  booktitle={Proceedings of the 2022 ACM Conference on Fairness, Accountability, and Transparency},
  pages={2035--2046},
  year={2022}
}

@article{Rando2022,
  author       = {Rando, A. and et al.},
  title        = {Red-Teaming the Stable Diffusion Safety Filter},
  journal      = {arXiv preprint arXiv:2210.04610},
  year         = {2022},
}

@inproceedings{Rombach2021,
  author       = {Rombach, R. and et al.},
  title        = {High-Resolution Image Synthesis with Latent Diffusion Models},
  booktitle    = {arXiv preprint arXiv:2112.10752},
  year         = {2021},
}

@misc{Shani2021,
  author       = {Shani, S. and et al.},
  title        = {Deepfakes and the Unsolved Challenge of Safety},
  year         = {2021},
  howpublished = {Retrieved from \url{https://www.dhs.gov/sites/default/files/publications/increasing_threats_of_deepfake_identities_0.pdf}}
}

@article{xu2017,
  author       = {Xu, T. and Zhang, P. and Huang, Q. and Zhang, H. and Gan, Z. and Huang, X. and He, X.},
  title        = {AttnGAN: Fine-Grained Text to Image Generation with Attentional Generative Adversarial Networks},
  journal      = {arXiv preprint arXiv:1711.10485},
  year         = {2017},
}

@article{zhu2019,
  author       = {Zhu, M. and Pan, P. and Chen, W. and Yang, Y.},
  title        = {DM-GAN: Dynamic Memory Generative Adversarial Networks for Text-to-Image Synthesis},
  journal      = {arXiv preprint arXiv:1904.01310},
  year         = {2019},
}

@article{crowson2022,
  author       = {Crowson, K. and Biderman, S.},
  title        = {VQGAN-CLIP: Open Domain Image Generation and Editing with Natural Language Guidance},
  journal      = {arXiv preprint arXiv:2204.08583},
  year         = {2022},
}

@misc{steinbruck2022,
  author       = {Steinbrück, A.},
  title        = {VQGAN+CLIP — How does it work?},
  year         = {2022},
  howpublished = {Retrieved from \url{https://alexasteinbruck.medium.com/vqgan-clip-how-does-it-work-210a5dca5e52}},
}

@article{ramesh2021,
  author       = {Ramesh, A. and Pavlov, M. and Goh, G. and Gray, S. and Voss, C. and Radford, A. and Chen, M. and Sutskever, I.},
  title        = {Zero-Shot Text-to-Image Generation},
  journal      = {arXiv preprint arXiv:2102.12092},
  year         = {2021},
}

@article{ramesh2022dalle2,
  author       = {Ramesh, A. and Dhariwal, P. and Nichol, A. and Chu, C. and Chen, M.},
  title        = {Hierarchical Text-Conditional Image Generation with CLIP Latents},
  journal      = {arXiv preprint arXiv:2204.06125},
  year         = {2022},
}

@article{runway2025,
    author = {Runway},
    title = {Runway Gen-4},
    journal = {Runway},
    year = {2025},
    howpublished = {Retrieved from \url{https://runwayml.com/research/introducing-runway-gen-4}}
}

@article{openai2024,
    author = {OpenAI},
    title = {Sora},
    journal = {OpenAI},
    year = {2024},
    howpublished = {Retrieved from \url{https://openai.com/sora/#features}}
}

@article{solaiman2023gradient,
    author = {Irene Solaiman},
    title = {The Gradient of Generative AI Release: Methods and Considerations},
    journal = {Arxiv},
    year = {2023},
    howpublished = {Retrieved from \url{https://arxiv.org/pdf/2302.04844}}
}

@inproceedings{umbach2024non,
  title={Non-consensual synthetic intimate imagery: Prevalence, attitudes, and knowledge in 10 countries},
  author={Umbach, Rebecca and Henry, Nicola and Beard, Gemma Faye and Berryessa, Colleen M},
  booktitle={Proceedings of the 2024 CHI Conference on Human Factors in Computing Systems},
  pages={1--20},
  year={2024}
}

@article{vincent2022stablediffusion,
  author  = {James Vincent},
  title   = {Stable Diffusion 2.0 makes it harder to mimic some artists' styles or generate photorealistic pictures of celebrities},
  journal = {The Verge},
  year    = {2022},
  month   = {November},
  day     = {24},
  url     = {https://www.theverge.com/2022/11/24/23476622/ai-image-generator-stable-diffusion-version-2-nsfw-artists-data-changes}
}

@article{ding2026stop,
  title={How to Stop Playing Whack-a-Mole: Mapping the Ecosystem of Technologies Facilitating AI-Generated Non-Consensual Intimate Images},
  author={Ding, Michelle L and Suresh, Harini and Venkatasubramanian, Suresh},
  journal={arXiv preprint arXiv:2602.04759},
  year={2026}
}

@inproceedings{podell2024sdxl,
  title={Sdxl: Improving latent diffusion models for high-resolution image synthesis},
  author={Podell, Dustin and English, Zion and Lacey, Kyle and Blattmann, Andreas and Dockhorn, Tim and M{\"u}ller, Jonas and Penna, Joe and Rombach, Robin},
  booktitle={International Conference on Learning Representations},
  volume={2024},
  pages={1862--1874},
  year={2024}
}

@techreport{securityhero2023,
  author       = {{Security Hero}},
  title        = {2023 State of Deepfakes: Realities, Threats, and Impact},
  year         = {2023},
  institution = {Security Hero},
  howpublished = {Retrieved from \url{https://www.securityhero.io/state-of-deepfakes/}},
}

@techreport{activefence2023,
    author = {ActiveFence},
    title = {The AI Surge in NCII Production},
    institution = {ActiveFence},
    year = {2023},
    howpublished = {Retrieved from \url{https://go.activefence.com/activefence-insights_the-ai-surge-in-ncii-production?utm_medium=email&_hsenc=p2ANqtz-8yla3mKudtDlQpJnTjOI0WrB_LXONfZbT-hL2GVZnRa3Di1Wftm52uW6SfHv4fyROKyGFzNdlp1K28FQzk5MkaaBwaDw&_hsmi=112628021&utm_content=112628021&utm_source=hs_automation}}
}

@article{wu2024finetunebench,
  title={FineTuneBench: How well do commercial fine-tuning APIs infuse knowledge into LLMs?},
  author={Wu, Eric and Wu, Kevin and Zou, James},
  journal={arXiv preprint arXiv:2411.05059},
  year={2024}
}

@misc{ujkani2025,
  author    = {Mir Ujkani},
  title     = {AI Product Validation With Beta Testing},
  year      = {2025},
  month     = apr,
  day       = {8},
  url       = {https://blog.betatesting.com/2025/04/08/ai-product-validation-with-beta-testing/},
  publisher = {BetaTesting Blog}
}

@article{cretu2025evaluating,
  title={Evaluating Concept Filtering Defenses against Child Sexual Abuse Material Generation by Text-to-Image Models},
  author={Cretu, Ana-Maria and Kireev, Klim and Abdalla, Amro and Obinna, Wisdom and Meier, Raphael and Bargal, Sarah Adel and Redmiles, Elissa M and Troncoso, Carmela},
  journal={arXiv preprint arXiv:2512.05707},
  year={2025}
}

@article{dugan2003regulating,
  title={Regulating What's Not Real: Federal Regulation in the Aftermath of Ashcroft v. Free Speech Coalition},
  author={Dugan, Kate},
  journal={. Louis ULJ},
  volume={48},
  pages={1063},
  year={2003},
  publisher={HeinOnline}
}

@article{krkic2025cultural,
  title={Cultural perspectives on AI usage and regulation in deepfake creation: How culture shapes AI practices},
  author={Krki{\'c}, Marijana},
  journal={International Communication of Chinese Culture},
  volume={12},
  number={2},
  pages={225--237},
  year={2025},
  publisher={Springer}
}

@misc{iwf2026ai,
  author = {{Internet Watch Foundation}},
  title = {{AI} becoming 'child sexual abuse machine' adding to 'dangerous' record levels of online abuse, {IWF} warns},
  year = {2026},
  month = jan,
  day = {16},
  howpublished = {\url{https://www.iwf.org.uk/news-media/news/ai-becoming-child-sexual-abuse-machine-adding-to-dangerous-record-levels-of-online-abuse-iwf-warns/}}
}

@article{iwf2024,
    author = {{Internet Watch Foundation}},
    title = {AI CSAM Report Update},
    year = 2024,
    journal = {IWF},
    url = {Retrieved from \url{https://admin.iwf.org.uk/media/nadlcb1z/iwf-ai-csam-report_update-public-jul24v13.pdf}},
}

@article{fusionxcivitai2025,
    author = {Anonymous User},
    title = {FusionX Ingredients Workflow},
    journal = {CivitAI},
    year = {2025},
    howpublished = {Retrieved from \url{https://civitai.com/models/1690979/fusionxingredientsworkflows}}
}

@article{thiel2023,
    author = {David Thiel},
    title = {Identifying and Eliminating CSAM in Generative ML Training Data and Models},
    journal = {retrieved from \url{https://purl.stanford.edu/kh752sm9123}},
    year = {2023},
}

@techreport{dalle2modelcard,
    author = {OpenAI},
    title = {DALL·E 2 Preview - Risks and Limitations},
    institution = OpenAI,
    year = 2022
}

@article{wan2025wan,
  title={Wan: Open and advanced Large-scale Video Generative Models},
  author={Wan, Team and Wang, Ang and Ai, Baole and Wen, Bin and Mao, Chaojie and Xie, Chen-Wei and Chen, Di and Yu, Feiwu and Zhao, Haiming and Yang, Jianxiao and others},
  journal={arXiv preprint arXiv:2503.20314},
  year={2025}
}

@article{kong2412hunyuanvideo,
  title={Hunyuanvideo: A systematic framework for large video generative models, 2025},
  author={Kong, W and Tian, Q and Zhang, Z and Min, R and Dai, Z and Zhou, J and Xiong, J and Li, X and Wu, B and Zhang, J and others},
  year = {2025},
  journal={URL https://arxiv. org/abs/2412.03603}
}

@techreport{NIST_ID012_2024,
  title        = {Comments on AI Executive Order Request for Information},
  author       = {{Thorn} and {ATIH}},
  institution  = {National Institute of Standards and Technology},
  type         = {Public Comment (ID012)},
  number       = {ID012},
  year         = {2024},
  month        = feb,
  day          = {1},
  url          = {https://www.nist.gov/system/files/documents/2024/02/15/ID012%20-%202024-02-01%2C%20Thorn%20and%20ATIH%2C%20Comments%20on%20AI%20EO%20RFI.pdf},
  note         = {Submitted in response to NIST’s RFI under Executive Order 14110}
}

@techreport{iwf2023ai-csam,
  title        = {How AI is Being Abused to Create Child Sexual Abuse Imagery},
  author       = {{Internet Watch Foundation (IWF)}},
  institution  = {Internet Watch Foundation},
  address      = {United Kingdom},
  year         = {2023},
  month        = {October},
  type         = {Technical report (Public version)},
  note         = {Contains descriptions of methods used to generate AI CSAM, along with verbatim perpetrator comments. Does not include AI CSAM images.},
  url          = {https://www.iwf.org.uk/media/q4zll2ya/iwf-ai-csam-report_public-oct23v1.pdf}
}

@article{birhane2311into,
  title={Into the LAIONs Den: Investigating Hate in Multimodal Datasets, November 2023},
  author={Birhane, Abeba and Prabhu, Vinay and Han, Sang and Boddeti, Vishnu Naresh and Luccioni, Alexandra Sasha},
  journal={URL http://arxiv. org/abs/2311.03449},
  year = {2023}
}

@article{lu2025concepts,
  title={When Are Concepts Erased From Diffusion Models?},
  author={Lu, Kevin and Kriplani, Nicky and Gandikota, Rohit and Pham, Minh and Bau, David and Hegde, Chinmay and Cohen, Niv},
  journal={arXiv preprint arXiv:2505.17013},
  year={2025}
}

@article{schneider2024image,
  title={When Image Generation Goes Wrong: A Safety Analysis of Stable Diffusion Models},
  author={Schneider, Matthias and Hagendorff, Thilo},
  journal={arXiv preprint arXiv:2411.15516},
  year={2024}
}

@article{zhang2025packing,
  title={Packing input frame context in next-frame prediction models for video generation},
  author={Zhang, Lvmin and Agrawala, Maneesh},
  journal={arXiv preprint arXiv:2504.12626},
  year={2025}
}

@article{casper2025open,
title={Open Technical Problems in Open-Weight AI Model Risk Management},
author={Casper, Stephen and O'Brien, Kyle and Longpre, Shayne and Seger, Elizabeth and Klyman, Kevin and Bommasani, Rishi and Nrusimha, Aniruddha and Shumailov, Ilia and Mindermann, S{\"o}ren and Basart, Steven and Rudzicz, Frank and Pelrine, Kellin and Ghosh, Avijit and Strait, Andrew and Kirk, Robert and Hendrycks, Dan and Henderson, Peter and Kolter, Zico and Irving, Geoffrey and Gal, Yarin and Bengio, Yoshua and Hadfield-Menell, Dylan},
year={2025},
}

@misc{simonovich2024nytheon,
  author = {Simonovich, Vitaly},
  title = {Cato CTRL Nytheon AI: A New Platform of Uncensored LLMs},
  year = {2024},
  month = {June},
  howpublished = {Cato Networks Blog},
  url = {https://www.catonetworks.com/blog/cato-ctrl-nytheon-ai-a-new-platform-of-uncensored-llms/},
  note = {Accessed: 2025-10-30}
}

@article{smith2023piracy,
  title={What the Online Piracy Data Tells Us About Copyright Policymaking},
  author={Smith, Michael},
  year={2023},
  journal={Hudson Institute},
  note={Studies show 11-14\% increase in legal sales when piracy links removed}
}

@article{danaher2017,
  title={Copyright Enforcement in the Digital Age},
  author = {Danaher, Brett et. al.},
  journal={Communications of the ACM},
  year={2017},
  note={Search behavior studies show effectiveness of demoting piracy sites}
}

@article{o2025deep,
  title={Deep Ignorance: Filtering Pretraining Data Builds Tamper-Resistant Safeguards into Open-Weight LLMs},
  author={O'Brien, Kyle and Casper, Stephen and Anthony, Quentin and Korbak, Tomek and Kirk, Robert and Davies, Xander and Mishra, Ishan and Irving, Geoffrey and Gal, Yarin and Biderman, Stella},
  journal={arXiv preprint arXiv:2508.06601},
  year={2025}
}

@article{williams2025regulating,
  title={On Regulating Downstream AI Developers},
  author={Williams, Sophie and Schuett, Jonas and Anderljung, Markus},
  journal={arXiv preprint arXiv:2503.11922},
  year={2025}
}

@article{petsiuk2022human,
  title={Human evaluation of text-to-image models on a multi-task benchmark},
  author={Petsiuk, Vitali and Siemenn, Alexander E and Surbehera, Saisamrit and Chin, Zad and Tyser, Keith and Hunter, Gregory and Raghavan, Arvind and Hicke, Yann and Plummer, Bryan A and Kerret, Ori and others},
  journal={arXiv preprint arXiv:2211.12112},
  year={2022}
}

@article{nichol2021glide,
  title={Glide: Towards photorealistic image generation and editing with text-guided diffusion models},
  author={Nichol, Alex and Dhariwal, Prafulla and Ramesh, Aditya and Shyam, Pranav and Mishkin, Pamela and McGrew, Bob and Sutskever, Ilya and Chen, Mark},
  journal={arXiv preprint arXiv:2112.10741},
  year={2021}
}

@book{henry2020image,
  title={Image-based sexual abuse: A study on the causes and consequences of non-consensual nude or sexual imagery},
  author={Henry, Nicola and McGlynn, Clare and Flynn, Asher and Johnson, Kelly and Powell, Anastasia and Scott, Adrian J},
  year={2020},
  publisher={Routledge}
}

@article{gieseke2020new,
  title={" The new weapon of choice": Law's current inability to properly address deepfake pornography},
  author={Gieseke, Anne Pechenik},
  journal={Vand. L. Rev.},
  volume={73},
  pages={1479},
  year={2020},
  publisher={HeinOnline}
}

@article{eggestein2014fighting,
  title={Fighting child pornography: A review of legal and technological developments},
  author={Eggestein, Jasmine V and Knapp, Kenneth J},
  journal={Journal of Digital Forensics, Security and Law},
  volume={9},
  number={4},
  pages={3},
  year={2014}
}

@inproceedings{yang2024sneakyprompt,
  title={Sneakyprompt: Jailbreaking text-to-image generative models},
  author={Yang, Yuchen and Hui, Bo and Yuan, Haolin and Gong, Neil and Cao, Yinzhi},
  booktitle={2024 IEEE symposium on security and privacy (SP)},
  pages={897--912},
  year={2024},
  organization={IEEE}
}

@misc{stable_diffusion_launch_2022,
  title        = {Stable Diffusion Launch Announcement},
  author       = {Joshua Lopez},
  year         = {2022},
  month        = aug,
  day          = {10},
  howpublished = {\url{https://stability.ai/news/stable-diffusion-announcement}},
  note         = {Stable Diffusion first stage release; model weights hosted via Hugging Face; co-led by Runway, LMU Munich / CompVis, etc.},
}

@article{sharma2024unlearning,
  title={Unlearning or concealment? a critical analysis and evaluation metrics for unlearning in diffusion models},
  author={Sharma, Aakash Sen and Sarkar, Niladri and Chundawat, Vikram and Mali, Ankur A and Mandal, Murari},
  journal={arXiv preprint arXiv:2409.05668},
  year={2024}
}

@inproceedings{wu2025unlearning,
  title={Unlearning concepts in diffusion model via concept domain correction and concept preserving gradient},
  author={Wu, Yongliang and Zhou, Shiji and Yang, Mingzhuo and Wang, Lianzhe and Chang, Heng and Zhu, Wenbo and Hu, Xinting and Zhou, Xiao and Yang, Xu},
  booktitle={Proceedings of the AAAI Conference on Artificial Intelligence},
  volume={39},
  number={8},
  pages={8496--8504},
  year={2025}
}

@article{zhang2024defensive,
  title={Defensive unlearning with adversarial training for robust concept erasure in diffusion models},
  author={Zhang, Yimeng and Chen, Xin and Jia, Jinghan and Zhang, Yihua and Fan, Chongyu and Liu, Jiancheng and Hong, Mingyi and Ding, Ke and Liu, Sijia},
  journal={Advances in neural information processing systems},
  volume={37},
  pages={36748--36776},
  year={2024}
}

@article{fuchi2024erasing,
  title={Erasing concepts from text-to-image diffusion models with few-shot unlearning},
  author={Fuchi, Masane and Takagi, Tomohiro},
  journal={arXiv preprint arXiv:2405.07288},
  volume={2},
  pages={1},
  year={2024}
}

@inproceedings{wu2025erasing,
  title={Erasing undesirable influence in diffusion models},
  author={Wu, Jing and Le, Trung and Hayat, Munawar and Harandi, Mehrtash},
  booktitle={Proceedings of the Computer Vision and Pattern Recognition Conference},
  pages={28263--28273},
  year={2025}
}

@article{yu2025forgetme,
  title={ForgetMe: Evaluating Selective Forgetting in Generative Models},
  author={Yu, Zhenyu and Idris, Mohd Yamani Inda and Wang, Pei},
  journal={arXiv preprint arXiv:2504.12574},
  year={2025}
}

@article{li2025towards,
  title={Towards Resilient Safety-driven Unlearning for Diffusion Models against Downstream Fine-tuning},
  author={Li, Boheng and Gu, Renjie and Wang, Junjie and Qi, Leyi and Li, Yiming and Wang, Run and Qin, Zhan and Zhang, Tianwei},
  journal={arXiv preprint arXiv:2507.16302},
  year={2025}
}

@article{gao2024meta,
  title={Meta-unlearning on diffusion models: Preventing relearning unlearned concepts},
  author={Gao, Hongcheng and Pang, Tianyu and Du, Chao and Hu, Taihang and Deng, Zhijie and Lin, Min},
  journal={arXiv preprint arXiv:2410.12777},
  year={2024}
}

@inproceedings{george2025illusion,
  title={The Illusion of Unlearning: The Unstable Nature of Machine Unlearning in Text-to-Image Diffusion Models},
  author={George, Naveen and Dasaraju, Karthik Nandan and Chittepu, Rutheesh Reddy and Mopuri, Konda Reddy},
  booktitle={Proceedings of the Computer Vision and Pattern Recognition Conference},
  pages={13393--13402},
  year={2025}
}

@inproceedings{suriyakumar2024unstable,
  title={Unstable unlearning: The hidden risk of concept resurgence in diffusion models},
  author={Suriyakumar, Vinith Menon and Alur, Rohan and Sekhari, Ayush and Raghavan, Manish and Wilson, Ashia C},
  booktitle={ICLR 2025 Workshop on Navigating and Addressing Data Problems for Foundation Models},
  year={2024}
}

@article{liu2024unlearning,
  title={Unlearning Concepts from Text-to-Video Diffusion Models},
  author={Liu, Shiqi and Tan, Yihua},
  journal={arXiv preprint arXiv:2407.14209},
  year={2024}
}

@article{zhu2024choose,
  title={Choose your anchor wisely: Effective unlearning diffusion models via concept reconditioning},
  author={Zhu, Jingyu and Zhang, Ruiqi and Lin, Licong and Mei, Song},
  year={2024}
}

@article{zhang2025concept,
  title={Concept Unlearning by Modeling Key Steps of Diffusion Process},
  author={Zhang, Chaoshuo and Lin, Chenhao and Zhao, Zhengyu and Yang, Le and Wang, Qian and Shen, Chao},
  journal={arXiv preprint arXiv:2507.06526},
  year={2025}
}

@inproceedings{geeng2025co,
  title={Co-Constructing the Future of Digital Intimacy},
  author={Geeng, Chris and Qin, Lucy and McDonald, Allison and Batool, Amna and Freed, Diana and Haimson, Oliver L and Hutson, Jevan and Redmiles, Elissa M and Stardust, Zahra and Wei, Miranda and others},
  booktitle={Companion Publication of the 2025 Conference on Computer-Supported Cooperative Work and Social Computing},
  pages={144--149},
  year={2025}
}

@inproceedings{zhang2024generate,
  title={To generate or not? safety-driven unlearned diffusion models are still easy to generate unsafe images... for now},
  author={Zhang, Yimeng and Jia, Jinghan and Chen, Xin and Chen, Aochuan and Zhang, Yihua and Liu, Jiancheng and Ding, Ke and Liu, Sijia},
  booktitle={European Conference on Computer Vision},
  pages={385--403},
  year={2024},
  organization={Springer}
}

@inproceedings{casper2024black,
  title={Black-box access is insufficient for rigorous ai audits},
  author={Casper, Stephen and Ezell, Carson and Siegmann, Charlotte and Kolt, Noam and Curtis, Taylor Lynn and Bucknall, Benjamin and Haupt, Andreas and Wei, Kevin and Scheurer, J{\'e}r{\'e}my and Hobbhahn, Marius and others},
  booktitle={Proceedings of the 2024 ACM Conference on Fairness, Accountability, and Transparency},
  pages={2254--2272},
  year={2024}
}

@article{che2025model,
  title={Model tampering attacks enable more rigorous evaluations of llm capabilities},
  author={Che, Zora and Casper, Stephen and Kirk, Robert and Satheesh, Anirudh and Slocum, Stewart and McKinney, Lev E and Gandikota, Rohit and Ewart, Aidan and Rosati, Domenic and Wu, Zichu and others},
  journal={arXiv preprint arXiv:2502.05209},
  year={2025}
}

@article{wallace2025estimating,
  title={Estimating worst-case frontier risks of open-weight llms},
  author={Wallace, Eric and Watkins, Olivia and Wang, Miles and Chen, Kai and Koch, Chris},
  journal={arXiv preprint arXiv:2508.03153},
  year={2025}
}

@phdthesis{monaghan2017impact,
  title={The impact of Self-Generated Images in online pornography},
  author={Monaghan, Andy},
  year={2017},
  school={Middlesex University}
}

@article{kobriger2021out,
  title={Out of our depth with deep fakes: How the law fails victims of deep fake nonconsensual pornography},
  author={Kobriger, Kate and Zhang, Janet and Quijano, Andrew and Guo, Joyce},
  journal={Rich. JL \& Tech.},
  volume={28},
  pages={204},
  year={2021},
  publisher={HeinOnline}
}

@article{wagner2019word,
  title={“The word real is no longer real”: Deepfakes, gender, and the challenges of ai-altered video},
  author={Wagner, Travis L and Blewer, Ashley},
  journal={Open Information Science},
  volume={3},
  number={1},
  pages={32--46},
  year={2019},
  publisher={De Gruyter Open}
}

@article{winter2020deepfakes,
  title={DeepFakes: uncovering hardcore open source on GitHub},
  author={Winter, Rachel and Salter, Anastasia},
  journal={Porn Studies},
  volume={7},
  number={4},
  pages={382--397},
  year={2020},
  publisher={Taylor \& Francis}
}

@article{williams2025there,
  title={“There Are No Limits!”: AI Undressing Apps and the Normalization of Nonconsensual Intimate Deepfakes},
  author={Williams, Kaylee},
  journal={Violence Against Women},
  pages={10778012251397966},
  year={2025},
  publisher={SAGE Publications Sage CA: Los Angeles, CA}
}

@article{arcila2025ai,
  title={AI Liability Along the Value Chain: Lessons from the Liability of Suppliers of Components in Product Liability Law},
  author={Arcila, Beatriz Botero},
  journal={Journal of Tort Law},
  volume={18},
  number={1},
  pages={247--281},
  year={2025},
  publisher={De Gruyter}
}

@techreport{IWF2026,
  author      = {{Internet Watch Foundation}},
  title       = {Harm Without Limits: {AI} Child Sexual Abuse Material Through the Eyes of Our Analysts},
  institution = {Internet Watch Foundation},
  year        = {2026},
  url         = {https://www.iwf.org.uk/about-us/why-we-exist/our-research/how-ai-is-being-abused-to-create-child-sexual-abuse-imagery/}
}

@article{louk2026deepfakes,
  title={Deepfakes, Real Enforcement Challenges},
  author={Louk, David S},
  journal={The Columbia Journal of Law \& the Arts},
  volume={49},
  number={4},
  pages={817--842},
  year={2026}
}

@article{abdalla2025gift,
  title={GIFT: Gradient-aware Immunization of diffusion models against malicious Fine-Tuning with safe concepts retention},
  author={Abdalla, Amro and Shaheen, Ismail and DeGenaro, Dan and Mallick, Rupayan and Raita, Bogdan and Bargal, Sarah Adel},
  journal={arXiv preprint arXiv:2507.13598},
  year={2025}
}

@misc{hashim2025,
  title={Why is Hugging Face hosting tools to make deepfake porn of teenage celebrities?},
  author={Shakeel Hashim},
  url={\url{https://www.transformernews.ai/p/why-ethical-ai-company--hugging-face-hosting-pornography-celebrities}},
  year={2025}
}

@misc{hfcontentpolicy2025,
  title        = {Content Policy},
  author       = {{Hugging Face}},
  year         = {2025},
  month        = April,
  howpublished = {\url{https://huggingface.co/content-policy}},
  note         = {Accessed: 2025-09-28}
}

@techreport{graphika2023,
    author = {Lakatos, Santiago},
    title = {A Revealing Picture},
    institution = {Graphika},
    year = {2023},
    howpublished = {Retrieved from \url{https://graphika.com/reports/a-revealing-picture}}

}

@misc{europol2025operation,
  title = {25 Arrested in Global Hit Against {AI}-Generated Child Sexual Abuse Material},
  author = {{Europol}},
  year = {2025},
  month = feb,
  day = {28},
  howpublished = {Press release},
  url = {https://www.europol.europa.eu/media-press/newsroom/news/25-arrested-in-global-hit-against-ai-generated-child-sexual-abuse-material},
  note = {Accessed: 2025-10-25}
}

@inproceedings{gibson2025analyzing,
  title={Analyzing the $\{$AI$\}$ Nudification Application Ecosystem},
  author={Gibson, Cassidy and Olszewski, Daniel and Brigham, Natalie Grace and Crowder, Anna and Butler, Kevin RB and Traynor, Patrick and Redmiles, Elissa M and Kohno, Tadayoshi},
  booktitle={34th USENIX Security Symposium (USENIX Security 25)},
  pages={1--20},
  year={2025}
}

@article{maini2025safety,
  title={Safety pretraining: Toward the next generation of safe ai},
  author={Maini, Pratyush and Goyal, Sachin and Sam, Dylan and Robey, Alex and Savani, Yash and Jiang, Yiding and Zou, Andy and Lipton, Zacharcy C and Kolter, J Zico},
  journal={arXiv preprint arXiv:2504.16980},
  year={2025}
}

@article{wagner2025perpetuating,
  title={Perpetuating Misogyny with Generative AI: How Model Personalization Normalizes Gendered Harm},
  author={Wagner, Laura and Cetinic, Eva},
  journal={arXiv preprint arXiv:2505.04600},
  year={2025}
}

@online{civitai2024transparency,
  title        = {Civitai 2024 Transparency Report},
  author       = {{Civitai}},
  year         = {2024},
  month        = {12},
  url          = {https://civitai.com/articles/10372/civitai-2024-transparency-report},
  urldate      = {2024-11-22},
  organization = {Civitai},
  note         = {Annual transparency report covering platform growth, financial metrics, and operational efforts}
}

@article{lefkowitz2026grok,
  title={GROK IMAGE GENERATION GOVERNANCE AUDIT Targeted Sexualization on X: An Independent Research Report},
  author={Lefkowitz, Dana},
  journal={Available at SSRN 6123306},
  year={2026}
}

@techreport{CCDH2026grok,
  author      = {{Center for Countering Digital Hate}},
  title       = {Grok Floods {X} with Sexualized Images of Women and Children},
  institution = {Center for Countering Digital Hate},
  year        = {2026},
  month       = {January},
  day         = {22},
  url         = {https://counterhate.com/research/grok-floods-x-with-sexualized-images/}
}

@inproceedings{wei2024exploring,
  title={Exploring the use of abusive generative AI models on Civitai},
  author={Wei, Yiluo and Zhu, Yiming and Hui, Pan and Tyson, Gareth},
  booktitle={Proceedings of the 32nd ACM International Conference on Multimedia},
  pages={6949--6958},
  year={2024}
}

@article{gorwa2024moderating,
  title={Moderating model marketplaces: Platform governance puzzles for AI intermediaries},
  author={Gorwa, Robert and Veale, Michael},
  journal={Law, Innovation and Technology},
  volume={16},
  number={2},
  pages={341--391},
  year={2024},
  publisher={Taylor \& Francis}
}

@inproceedings{hawkins2025deepfakes,
  title={Deepfakes on Demand: The rise of accessible non-consensual deepfake image generators: The rise of accessible non-consensual deepfake image generators},
  author={Hawkins, Will and Mittelstadt, Brent and Russell, Chris},
  booktitle={Proceedings of the 2025 ACM Conference on Fairness, Accountability, and Transparency},
  pages={1602--1614},
  year={2025}
}

@article{laffier2023deepfakes,
  title={Deepfakes and harm to women},
  author={Laffier, Jennifer and Rehman, Aalyia},
  journal={Journal of Digital Life and Learning},
  volume={3},
  number={1},
  pages={1--21},
  year={2023}
}

@misc{laion2022laion5b,
  title        = {{LAION-5B: A New Era of Open Large-Scale Multi-Modal Datasets}},
  author       = {Beaumont, Romain and the LAION Team},
  year         = {2022},
  howpublished = {\url{https://laion.ai/blog/laion-5b/}},
  note         = {Accessed: 2025-10-02}
}

@misc{laion2024relaion5b,
  author       = {LAION},
  title        = {Releasing Re-LAION-5B: transparent iteration on LAION-5B with additional safety fixes},
  year         = {2024},
  howpublished = {\url{https://laion.ai/blog/relaion-5b/}},
  note         = {Accessed: 2025-10-02}
}

@article{Salvaggio2024,
  author       = {Eryk Salvaggio},
  title        = {LAION-5B, Stable Diffusion 1.5, and the Original Sin of Generative AI},
  journal      = {TechPolicy.Press},
  year         = {2024},
  url          = {https://www.techpolicy.press/laion5b-stable-diffusion-and-the-original-sin-of-generative-ai/},
}

@article{bhandari2025forecasting,
  title={Forecasting Open-Weight AI Model Growth on HuggingFace},
  author={Bhandari, Kushal Raj and Chen, Pin-Yu and Gao, Jianxi},
  journal={arXiv preprint arXiv:2502.15987},
  year={2025}
}

@article{bommasani_considerations_2024,
	title = {Considerations for governing open foundation models},
	volume = {386},
	url = {https://www.science.org/doi/abs/10.1126/science.adp1848},
	doi = {10.1126/science.adp1848},
	number = {6718},
	urldate = {2025-10-07},
	journal = {Science},
	author = {Bommasani, Rishi and Kapoor, Sayash and Klyman, Kevin and Longpre, Shayne and Ramaswami, Ashwin and Zhang, Daniel and Schaake, Marietje and Ho, Daniel E. and Narayanan, Arvind and Liang, Percy},
	month = oct,
	year = {2024},
	note = {Publisher: American Association for the Advancement of Science},
	pages = {151--153},
}

@misc{kapoor_societal_2024,
	title = {On the {Societal} {Impact} of {Open} {Foundation} {Models}},
	url = {http://arxiv.org/abs/2403.07918},
	doi = {10.48550/arXiv.2403.07918},
	urldate = {2025-10-07},
	publisher = {arXiv},
	author = {Kapoor, Sayash and Bommasani, Rishi and Klyman, Kevin and Longpre, Shayne and Ramaswami, Ashwin and Cihon, Peter and Hopkins, Aspen and Bankston, Kevin and Biderman, Stella and Bogen, Miranda and others},
	month = feb,
	year = {2024},
	note = {arXiv:2403.07918 [cs]},
}

@misc{francois_different_2025,
	title = {A {Different} {Approach} to {AI} {Safety}: {Proceedings} from the {Columbia} {Convening} on {Openness} in {Artificial} {Intelligence} and {AI} {Safety}},
	shorttitle = {A {Different} {Approach} to {AI} {Safety}},
	url = {http://arxiv.org/abs/2506.22183},
	doi = {10.48550/arXiv.2506.22183},
	urldate = {2025-10-07},
	publisher = {arXiv},
	author = {François, Camille and Péran, Ludovic and Bdeir, Ayah and Dziri, Nouha and Hawkins, Will and Jernite, Yacine and Kapoor, Sayash and Shen, Juliet and Khlaaf, Heidy and Klyman, Kevin and others},
	month = jun,
	year = {2025},
	note = {arXiv:2506.22183 [cs]},
}

@article{maiberg2025,
    title={Hugging Face Is Hosting 5,000 Nonconsensual AI Models of Real People},
    author={Maiberg, Emanuel},
    journal={404 Media},
    year={2025},
    url={\url{https://www.404media.co/hugging-face-is-hosting-5-000-nonconsensual-ai-models-of-real-people/}},
    note={Accessed: 2025-10-07}
}

@misc{genmo2025mochi1,
  author       = {{Genmo, Inc.}},
  title        = {Mochi 1: A New SOTA in Open-Source Video Generation Models},
  howpublished = {\url{https://www.genmo.ai/blog}},
  year         = {2025},
  note         = {Accessed: 2025-10-15}
}

@article{hacohen2024ltx,
  title={Ltx-video: Realtime video latent diffusion},
  author={HaCohen, Yoav and Chiprut, Nisan and Brazowski, Benny and Shalem, Daniel and Moshe, Dudu and Richardson, Eitan and Levin, Eran and Shiran, Guy and Zabari, Nir and Gordon, Ori and others},
  journal={arXiv preprint arXiv:2501.00103},
  year={2024}
}

@article{yang2024cogvideox,
  title={Cogvideox: Text-to-video diffusion models with an expert transformer},
  author={Yang, Zhuoyi and Teng, Jiayan and Zheng, Wendi and Ding, Ming and Huang, Shiyu and Xu, Jiazheng and Yang, Yuanming and Hong, Wenyi and Zhang, Xiaohan and Feng, Guanyu and others},
  journal={arXiv preprint arXiv:2408.06072},
  year={2024}
}

@article{blattmann2023stable,
  title={Stable video diffusion: Scaling latent video diffusion models to large datasets},
  author={Blattmann, Andreas and Dockhorn, Tim and Kulal, Sumith and Mendelevitch, Daniel and Kilian, Maciej and Lorenz, Dominik and Levi, Yam and English, Zion and Voleti, Vikram and Letts, Adam and others},
  journal={arXiv preprint arXiv:2311.15127},
  year={2023}
}

@article{zhou2025stable,
  title={Stable virtual camera: Generative view synthesis with diffusion models},
  author={Zhou, Jensen Jinghao and Gao, Hang and Voleti, Vikram and Vasishta, Aaryaman and Yao, Chun-Han and Boss, Mark and Torr, Philip and Rupprecht, Christian and Jampani, Varun},
  journal={arXiv preprint arXiv:2503.14489},
  year={2025}
}

@article{agarwal2025cosmos,
  title={Cosmos world foundation model platform for physical ai},
  author={Agarwal, Niket and Ali, Arslan and Bala, Maciej and Balaji, Yogesh and Barker, Erik and Cai, Tiffany and Chattopadhyay, Prithvijit and Chen, Yongxin and Cui, Yin and Ding, Yifan and others},
  journal={arXiv preprint arXiv:2501.03575},
  year={2025}
}

@inproceedings{wang2025seedvr,
  title={Seedvr: Seeding infinity in diffusion transformer towards generic video restoration},
  author={Wang, Jianyi and Lin, Zhijie and Wei, Meng and Zhao, Yang and Yang, Ceyuan and Loy, Chen Change and Jiang, Lu},
  booktitle={Proceedings of the Computer Vision and Pattern Recognition Conference},
  pages={2161--2172},
  year={2025}
}

@article{lin2024animatediff,
  title={Animatediff-lightning: Cross-model diffusion distillation},
  author={Lin, Shanchuan and Yang, Xiao},
  journal={arXiv preprint arXiv:2403.12706},
  year={2024}
}

@misc{alexios_mantzarlis_ai_2025,
    title = {{AI} {Nudifiers} continue to reach millions and make millions},
    url = {https://indicator.media/p/ai-nudifiers-continue-to-reach-millions-and-make-millions},
    journal = {The Indicator},
    author = {{Alexios Mantzarlis} and {Santiago Lakatos}},
    month = jul,
    year = {2025},
}

@misc{rainn2025grok,
  title = {Grok's 'Spicy' {AI} Video Setting Will Lead to Sexual Abuse},
  author = {{RAINN}},
  year = {2025},
  month = aug,
  day = {28},
  howpublished = {\url{https://rainn.org/groks-spicy-ai-video-setting-will-lead-to-sexual-abuse/}},
  note = {Accessed: 2025-11-20},
  organization = {Rape, Abuse \& Incest National Network}
}

@misc{maiberg2025grok,
  title = {Elon {Musk's} {Grok} {AI} Will 'Remove Her Clothes' In Public, On {X}},
  author = {Maiberg, Emanuel},
  year = {2025},
  month = may,
  day = {6},
  howpublished = {\url{https://www.404media.co/elon-musks-grok-ai-will-remove-her-clothes-in-public-on-x/}},
  note = {Accessed: 2025-11-20},
  organization = {404 Media}
}

@misc{my_image_my_choice_deepfake_2024,
    title = {{DEEPFAKE} {ABUSE}:  {LANDSCAPE} {ANALYSIS}: {The} {Exponential} {Rise} of {Deepfake} {Abuse} in 2023 - 2024},
    url = {https://www.canva.com/design/DAGLHpt6WlY/Hfztqtw_-tKza_2l1cPNrA/view?utm_content=DAGLHpt6WlY&utm_campaign=designshare&utm_medium=link&utm_source=editor#1},
    author = {{My Image My Choice}},
    year = {2024},
}

@misc{sen_cruz_ted_r-tx_take_2023,
    title = {{TAKE} {IT} {DOWN} {Act}},
    url = {https://www.congress.gov/bill/118th-congress/senate-bill/4569/cosponsors},
    author = {{Sen. Cruz, Ted [R-TX]}},
    year = {2023},
}

@misc{patrick_trueman_esq_github_2023,
    title = {{GitHub} hosting source code for sexually exploitative technology, facilitating image-based sexual abuse ({IBSA}), sexual exploitation, and promoting the dangerous use of generative-{AI}},
    url = {https://endsexualexploitation.org/wp-content/uploads/Microsoft-GitHub-Notification-Letter_DDL-2023_FINAL.pdf},
    author = {{Patrick Trueman, Esq.} and {Dawn Hawkins}},
    month = apr,
    year = {2023},
}

@article{flynn_deepfakes_2022,
    title = {Deepfakes and {Digitally} {Altered} {Imagery} {Abuse}: {A} {Cross}-{Country} {Exploration} of an {Emerging} form of {Image}-{Based} {Sexual} {Abuse}},
    volume = {62},
    copyright = {https://academic.oup.com/pages/standard-publication-reuse-rights},
    issn = {0007-0955, 1464-3529},
    shorttitle = {Deepfakes and {Digitally} {Altered} {Imagery} {Abuse}},
    url = {https://academic.oup.com/bjc/article/62/6/1341/6448791},
    doi = {10.1093/bjc/azab111},
    language = {en},
    number = {6},
    urldate = {2025-04-02},
    journal = {The British Journal of Criminology},
    author = {Flynn, Asher and Powell, Anastasia and Scott, Adrian J and Cama, Elena},
    month = oct,
    year = {2022},
    pages = {1341--1358},
}

@article{collier2026grok,
  author = {Collier, Kevin and Goggin, Ben and Ingram, David and Horvath, Bruna},
  title = {Elon Musk's X limits some sexual deepfakes after backlash, but Grok will still make the images},
  journal = {NBC News},
  year = {2026},
  month = {January},
  day = {9},
  note = {Analysis by deepfake researcher Genevieve Oh found Grok produced 7,751 sexualized images in one hour on January 8, 2026},
  url = {https://www.nbcnews.com/tech/internet/x-paywall-ai-image-grok-app-bikini-allows-sexual-deepfakes-rcna252647}
}

@article{longpre2024responsible,
  title={The responsible foundation model development cheatsheet: A review of tools \& resources},
  author={Longpre, Shayne and Biderman, Stella and Albalak, Alon and Schoelkopf, Hailey and McDuff, Daniel and Kapoor, Sayash and Klyman, Kevin and Lo, Kyle and Ilharco, Gabriel and San, Nay and others},
  journal={arXiv preprint arXiv:2406.16746},
  year={2024}
}

@article{seger2024open,
  title={Open horizons: exploring nuanced technical and policy approaches to openness in AI},
  author={Seger, Elizabeth},
  year={2024},
  publisher={Demos}
}

@inproceedings{srikumar2024risk,
  title={Risk mitigation strategies for the open foundation model value chain},
  author={Srikumar, Madhulika and Chang, Jiyoo and Chmielinski, Kasia},
  booktitle={Insights from PAI workshop Co-hosted with github},
  year={2024}
}

@article{thorn2024safety,
  title={Safety by Design for Generative AI: Preventing Child Sexual Abuse},
  author={Thorn, All Tech Is Human},
  journal={Thorn, All Tech Is Human. https://info. thorn. org/hubfs/thorn-safety-by-design-for-generative-AI. pdf},
  year={2024}
}

@article{ghosh2026marketplace,
  title={A Marketplace for AI-Generated Adult Content and Deepfakes},
  author={Ghosh, Shalmoli and DeVerna, Matthew R and Menczer, Filippo},
  journal={arXiv preprint arXiv:2601.09117},
  year={2026}
}

@article{pang2024towards,
  title={Towards understanding unsafe video generation},
  author={Pang, Yan and Xiong, Aiping and Zhang, Yang and Wang, Tianhao},
  journal={arXiv preprint arXiv:2407.12581},
  year={2024}
}

@article{mink2026unlimited,
  title={" Unlimited Realm of Exploration and Experimentation": Methods and Motivations of AI-Generated Sexual Content Creators},
  author={Mink, Jaron and Qin, Lucy and Redmiles, Elissa M},
  journal={arXiv preprint arXiv:2601.21028},
  year={2026}
}

@article{li2026towards,
  title={Towards resilient safety-driven unlearning for diffusion models against downstream fine-tuning},
  author={Li, Boheng and Gu, Renjie and Wang, Junjie and Qi, Leyi and Li, Yiming and Wang, Run and Qin, Zhan and Zhang, Tianwei},
  journal={Advances in Neural Information Processing Systems},
  volume={38},
  pages={17754--17789},
  year={2026}
}

\clearpage

\appendix

\section{Ethical Considerations}

Our research involved analyzing publicly accessible content, including NSFW material. We adhered to the following ethical guidelines: (1) We did not seek, download, store, or view non-consensual intimate imagery, and (2) while we name platforms and models to enable policy action, we have withheld specific search terms, links, and website names that could drive traffic toward websites that may facilitate harm. 
% Our correspondence with developers (discussed in Section~\ref{sec:reporting}) was in their capacity as representatives of the company, not as individuals. We have not reported the identity of any individuals with whom we corresponded. Our correspondence was with companies (not individuals) and was about models (not humans). As such, this was not a form of human subjects research\footnote{\url{https://ori.hhs.gov/content/chapter-3-The-Protection-of-Human-Subjects-Definitions}}.

\section{Methodological Details} \label{app:details}

\textbf{Search for online platforms:} Using web searches, analysis of internet fora, and discussions in past literature, we identified over 50 websites, social media sites, apps, platforms, and Discord servers. We ultimately finalized on a selection of subreddits, Civitai, and a site dedicated to the archival of Civitai models because they uniquely allowed us to analyze the number of search hits or tags for SFW and NSFW video models or AI videos. The rest of the online platforms that we investigated did not allow for quantifying search hits for different models or model-generated videos. 

\textbf{Subreddit selection:} We selected 3 SFW and 3 NSFW subreddits dedicated to AI images and videos. For each category, we chose the most popular subreddits, all of which with over 100,000 members. The SFW subreddits that we selected were r/aiArt, r/StableDiffusion, r/aivideo. However, we omit the names of the NSFW subreddits to avoid driving internet traffic to them. However, readers can request the names with an email to us.

\textbf{Model identification search terms:} To identify instances of each video generation model across online platforms (Reddit, Civitai, and the Civitai archive), we used multiple search terms per model to account for different naming conventions, abbreviations, and variations used by the online community. These search terms were developed through preliminary exploration of how users reference these models in practice. For example, for Stable Video Diffusion, we searched for both the full name "stable video diffusion" and the common abbreviation "SVD." Similarly, for Wan 2.x, we used both "Wan" and "Wan2" to capture different naming patterns. The complete list of search terms used for each model is shown in Table~\ref{tab:search_terms}.

\begin{table}[htb]
\centering
\caption{\textbf{Search terms used to identify models in online platforms.} For each model, we used multiple search terms to capture different naming conventions and abbreviations used by the community.}
\label{tab:search_terms}
\begin{tabular}{@{}ll@{}}
\toprule
\textbf{Model} & \textbf{Search Terms} \\ \midrule
Wan 2.x & Wan, Wan2 \\
HunyuanVideo & Hunyuan, hunyuanvideo \\
Mochi 1 & mochi, mochi-1 \\
CogVideoX & CogVideoX \\
LTX-Video & LTXV, LTX-Video \\
Stable Video Diffusion & SVD, stable video diffusion \\
Stable Virtual Camera & SVC, stable virtual camera \\
AnimateDiff Lightning & AnimateDiff, AnimateDiff-Lightning \\
Cosmos & Cosmos \\
SeedVR2 & SeedVR2 \\
Stable Diffusion 1.x & SD 1, SD 1.4, SD 1.5, stable diffusion 1 \\
Stable Diffusion 2.x & SD 2, SD 2.1, stable diffusion 2 \\ \bottomrule
\end{tabular}
\end{table}

\textbf{Img2Vid and Framepacking tools:} In Section~\ref{sec:patterns}, we restrict our analysis to AI video generators. However, it is important to note that a substantial fraction of AI videos online come from image generators scaffolded using tools such as Img2Vid and Framepack \citep{zhang2025packing}. However, these techniques are generally more time-consuming and typically produce lower-quality video than modern video generators.

\textbf{The NSFW $\propto$ AIG-NCII assumption:} In Section~\ref{sec:patterns}, we assume in our analysis that the uses of a model for NSFW content correlate with its uses for AIG-NCII. This is principally due to the ethical and legal hazards of searching for online AIG-NCII as well as the underground nature with which this content is generally made and distributed. We note that a flaw with this assumption relates to how some NSFW videos generated by AI systems are in a cartoon/anime style. For example, some models, such as {PurpleSmart.AI's Pony} models\footnote{\url{https://purplesmart.ai/pony/content7x}}, specialize in non-realistic forms of character art generation. However, from our exploration of videos generated with the models in Section~\ref{sec:patterns}, none appear to be disproportionately used for non-realistic NSFW character art compared to photorealistic NSFW content.

% \section{Open Video Models Developer Transparency} \label{app:transparency}

% Table~\ref{tab:reporting} summarizes if and what the AI video developers from Section~\ref{sec:video} report related to risk management. Model developers usually do not discuss mitigations related to AIG-NCII or AIG-CSAM. And when mitigations are discussed, it is often with very little detail.

\end{document}